\documentclass[aps,pre,onecolumn,showpacs,superscriptaddress,groupedaddress,rmp]{revtex4-2}

\usepackage{titlesec}
\usepackage{etoolbox} % for patching

\renewcommand\thesection{\arabic{section}}
\renewcommand\thesubsection{\arabic{section}.\arabic{subsection}}

\titleformat{\section}[hang]{\bfseries}{\thesection.}{0.5em}{}{}
\titleformat{\subsection}[hang]{\bfseries}{\thesubsection.}{0.5em}{}{}

\makeatletter
\let\oldappendix\appendix
\renewcommand{\appendix}{%
  \oldappendix
  \renewcommand\thesection{\Alph{section}}
  \titleformat{\section}[hang]{\bfseries}{Appendix~\thesection:}{0.5em}{}{}
}
\makeatother

\usepackage{scrextend}
\usepackage[toc,page]{appendix}
\usepackage{verbatim}
\usepackage{graphicx}
\usepackage{dcolumn}
\usepackage{hyperref}
\usepackage{mathtools}
\usepackage{bm}
\usepackage{lipsum}
\usepackage{subcaption}
\usepackage{color}
\allowdisplaybreaks
\usepackage{amsmath}
\usepackage[justification=raggedright, singlelinecheck=false]{caption}
\usepackage{xcolor}
\usepackage[makeroom]{cancel} 

\hypersetup{
   colorlinks=true,
   linkcolor=blue,
   citecolor=blue,
   filecolor=blue,
   urlcolor=blue
}

\begin{document}

\title{Limitations to Chemotactic Concentration Sensing during $Ca^{2+}$ Signaling}
 
\author{Swoyam Srirupa }
\author{Pradeep}
 \author{Vaibhav Wasnik}
 \affiliation{School of Physical Sciences,Indian Institute of Technology,Goa, Zip Code}
 
\date{\today}

\begin{abstract} 
\textbf{Abstract.}
Living cells sense noisy biochemical signals crucial for survival, yet models incorporating intracellular signaling are limited.  This study examines how cells sense chemotactic concentrations through phosphorylation readouts in $Ca^{2+}$ signaling, which is ubiquitous in most eukaryotic cells.  {Using stochastic simulations and analytical calculations we find} that concentration sensing remains robust to variations in cytoplasmic reaction rates once they exceed a certain value, suggesting a potential evolutionary advantage that allows cells to optimize other signaling tasks without compromising concentration sensing accuracy. Our analysis demonstrates theoretically that  \textit{Dictyostelium} is capable of sensing very low concentrations of cyclic adenosine monophosphate (cAMP) as is experimentally seen.

\textbf{Keywords:} $Ca^{2+}$ signaling

\end{abstract}

\maketitle

\section{Introduction}

Living cells execute essential functions such as division, growth, differentiation \cite{Grandjean2004}, and apoptosis \cite{Nagata2018}. They also navigate chemical gradients through chemotaxis \cite{Berg1972, Haastert2007, Thar2003, Fuller2010} and engage in collective migration \cite{Siedlik2017}, crucial for embryonic development, wound healing, and cancer progression. These processes are guided by chemical signals like hormones, neurotransmitters, or other molecules. Accurate signal interpretation is vital, as errors can cause malfunctions, disease, or increased predation risk.

Cells excel at decoding signals, as evidenced by numerous experimental studies. Rod cell photoreceptors detect single photons \cite{Hecht1942, Rieke1998}. E. coli sense $3.2$ nM L-aspartate \cite{Mao2003}, about three molecules in the cell volume. Dictyostelium cells can detect a cAMP gradient with a concentration difference of $1-5\%$ across the cell diameter \cite{Haastert2007}.

These observations raise fundamental theoretical questions: What are the physical limits of measurement in cellular systems? Has evolution optimized cellular machinery to perform measurements approaching these limits? \cite{Berg1977} first studied the theoretical constraints on concentration measurement, showing that the chemotaxis machinery in \textit{E. coli} appears optimally designed. \cite{Endres2009Concentration} investigated concentration sensing using maximum likelihood estimation  on time series data of  receptor occupancy. The impact of ligand-receptor kinetics \cite{Bialek2005, Kaizu2014} and receptor cooperativity \cite{Bialek2008, Skoge2011, Skoge2013} on concentration sensing has also been studied.

Cells detect and respond to spatial chemical gradients \cite{Berg1972, Haastert2007, Fuller2010}. The impact of diffusion \cite{Endres2008}, ligand-receptor kinetics \cite{Endres2009Gradient}, and receptor cooperativity \cite{Hu2010} in gradient sensing have been studied.

However, these studies focus on measurement limits at the cell surface, neglecting intracellular signaling effects. \cite{Govern2012} demonstrated that non-uniform time averaging can surpass Berg-Purcell calculations. \cite{GovernDec2014, GovernNov2014} considered the problem of energy constraints. The accuracy of positional readout in calcium signaling was analyzed in \cite{Wasnik2019PRL, Wasnik2019PRE}, the influence of signaling cascades on concentration sensing in \cite{Biswal2024}, and the accuracy of glutamate readout in neuronal signaling in \cite{Biswal2023}.

%Our study aims to advance the understanding of how intracellular signaling networks influence the sensing of extracellular chemical concentrations   over   long measurement times.

Here, we study how accurately cells can sense chemotactic concentrations through protein phosphorylation readouts in the cytoplasm once cellular processes, including ligand-receptor kinetics and biochemical reactions involved in signal transduction, reach a steady state. Our approach integrates simulations with theoretical analysis, focusing on nonlinear signaling networks, particularly calcium signaling pathways. Calcium ions ($Ca^{2+}$) act as secondary messengers, while extracellular ligands function as primary messengers. These pathways regulate numerous biological processes, including fertilization, secretion, contraction, cell-cycle progression, proliferation, apoptosis, and cognitive functions such as learning and memory \cite{Cullen2002}.

Using stochastic simulations and approximate analytical solutions of coupled nonlinear rate equations by linearizing around the steady state, we show that concentration sensing remains robust against variations in cytoplasmic rate constants once they exceed a certain value, meaning the sensing error becomes insensitive to a particular cytoplasmic rate constant when it is sufficiently large compared to other rates in the system.

 %................theory...................

\section{Theory}

\begin{figure}
\centering
\includegraphics[width=0.6\textwidth]{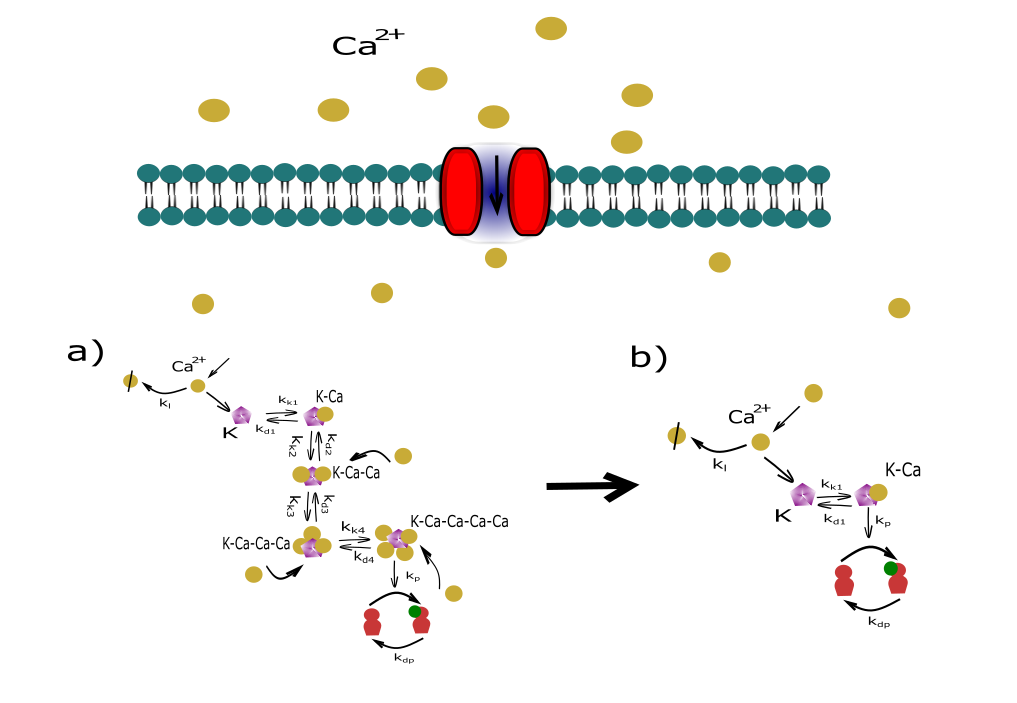}
\caption{The figure illustrates calcium signaling pathways. In panel (a), the activation of the kinase is triggered by the attachment of four calcium ions. However, this process introduces more noise into the system. To simplify the analysis, we have considered a model where the kinase is activated by the binding of a single calcium ion, as shown in panel (b).}  \label{singlecalcium}   
   \end{figure} 

$Ca^{2+}$ signaling is initiated by the binding of ligands to membrane-bound receptors that are permeable to $Ca^{2+}$. For example, in neurons, the binding of neurotransmitters to cell surface receptors such as NMDA and AMPA receptors, which are permeable to $Ca^{2+}$, triggers the opening of calcium channels \cite{Pchitskaya2018}. The opening of these channels facilitates the influx of $Ca^{2+}$ into the cytoplasm (see Fig.~\ref{singlecalcium}), leading to an increase in cytoplasmic $Ca^{2+}$ concentration. Within the cell, $Ca^{2+}$ binds to calmodulin (CaM), a calcium-binding protein. CaM is denoted by $K$ in Fig.~\ref{singlecalcium}. It contains four EF-hand motifs, each capable of binding a single $Ca^{2+}$ ion. Upon occupancy of all four binding sites, it undergoes a conformational change that enables it to interact with and activate downstream targets, such as $Ca^{2+}$-CaM dependent kinases (e.g., CaMKII) \cite{clapham}. Once activated, these kinases catalyze the phosphorylation of specific substrate proteins, thereby regulating vital cellular processes including muscle contraction, neurotransmitter release, and gene expression. Termination of $Ca^{2+}$ signaling requires the reduction of cytoplasmic free calcium levels, typically achieved by active transport of $Ca^{2+}$  out of the cytoplasm via pumps and exchangers. In our model, we consider a simplified version of this pathway in which $Ca^{2+}$ binds to a calmodulin (\( K \)) and resulting calcium-calmodulin complex directly phosphorylate target protein, thus bypassing the intermediate role of $Ca^{2+}$-CaM dependent kinases. The mechanism, involving the sequential binding of four $Ca^{2+}$ ions to the calmodulin, is illustrated in Fig.~\ref{singlecalcium}(a). However, to reduce complexity, we assume that calmodulin activation occurs upon the binding of a single \( \mathrm{Ca}^{2+} \) ion. This leads to a simplified model as depicted in Fig.~\ref{singlecalcium}(b). This simplification is made to estimate the lower bound of sensing accuracy. Including the full set of biochemical steps would increase the system's complexity and introduce additional sources of noise, which may, in turn, elevate the sensing error.

In the simplified model illustrated in Fig.~\ref{singlecalcium}(b), the cell is exposed to a static extracellular ligand concentration \( c \) at time $t=0$. Ligand molecules bind stochastically to cell surface receptors with a rate \( k_+ c \), and dissociate at a rate \( k_- \). At \( t = 0 \), the intracellular $Ca^{2+}$ concentration is assumed to be zero, and the concentration of unbound calmodulin (\( K \)) is denoted by \( C_{K0} \). Upon ligand binding, a stochastic influx of $Ca^{2+}$ is triggered. The influx rate is denoted by \( I_{Ca} \), while the efflux rate (representing the leaving of \( \mathrm{Ca}^{2+} \) from the cytoplasm) is given by \( k_l \). Within the cytoplasm, $Ca^{2+}$ binds to \( K \) at a rate \( k_{k1} \), forming a calmodulin-calcium complex \( (K\text{-}Ca) \), and unbinds at a rate \( k_{d1} \). The \( K\text{-}Ca \) complex catalyzes the phosphorylation of a target protein at a rate \( k_p \), initiating a downstream cellular response. The phosphorylated protein is subsequently dephosphorylated by a protein phosphatase enzyme (not shown in Fig.~\ref{singlecalcium}, assume it is in large quantity) at a rate \( k_{dp} \), rendering it inactive and available for future phosphorylation.

 {Let \( p(t) \) denote the probability that a receptor is occupied at time \( t \), with \( p(0) = 0 \), meaning the receptor is initially unoccupied. This corresponds to the moment when the cell is exposed to a static ligand concentration \( c \) at \( t = 0 \). Note that \( p(t) \) does not represent the instantaneous binary state of a single receptor, but rather the probability of occupancy across an ensemble or over time. Let \( C_{\mathrm{Ca}}(t) \), \( C_K(t) \), \( C_{K\text{-}Ca}(t) \), and \( C_{Pr}(t) \) denote the time-dependent concentrations of calcium ions, unbound calmodulin, the calmodulin–calcium complex, and phosphorylated protein, respectively. These species are initialized at \( t = 0 \) with the following conditions: \( C_{\mathrm{Ca}}(0) = 0 \), \( C_K(0) = C_{K0} \), \( C_{K\text{-}Ca}(0) = 0 \), and \( C_{Pr}(0) = 0 \), consistent with the model assumptions. The cell uses the instantaneous readout of phosphorylation level at measurement time $t=T$ as the output to infer the input ligand concentration \( c \), based on the input–output relationship \( \left( \frac{\Delta c_{\mathrm{rms}}}{\bar{c}} \right)^2 = \frac{\sigma_{C_{Pr}}^2}{\bar{c}^2 \left( \frac{d\bar{C}_{Pr}}{d\bar{c}} \right)^2} \) \cite{Bialek2008, Govern2012}, which quantifies the root-mean-square (RMS) error in estimating \( c \). To evaluate this expression, we require the variance \( \sigma_{C_{Pr}}^2 \), which is obtained by analyzing the time evolution of the system, governed by the following set of rate equations:}
 {
\begin{align}
    \frac{dp(t)}{dt} &= k_+ c \, (1 - p(t)) - k_- p(t) \\
    \frac{dC_{Ca}(t)}{dt} &= I_{Ca} \, p(t) -(k_l + k_{k1} C_K(t)) C_{Ca}(t) \nonumber \\ 
    & \quad + k_{d1} C_{K-Ca}(t) \\
    \frac{dC_{K}(t)}{dt} &= k_{d1} C_{K-Ca}(t) - k_{k1} C_K(t) \, C_{Ca}(t) \\
    \frac{dC_{K-Ca}(t)}{dt} &= k_{k1} C_K(t) \, C_{Ca}(t) - k_{d1} C_{K-Ca}(t) \\
    \frac{dC_{Pr}(t)}{dt} &= k_p C_{K-Ca}(t) - k_{dp} C_{Pr}(t).
\label{eq:rate_eqs}
\end{align}
}
 {Here, \( p(t) \) captures the extrinsic fluctuations arising from stochastic ligand–receptor binding and unbinding, while the remainder of the system is treated deterministically. This approach allows us to focus on how extrinsic fluctuations propagate through the downstream signaling network. We ignore intrinsic fluctuations downstream of the receptor for two main reasons. First, the concentrations (or molecule numbers) of downstream species are typically high (e.g., on the order of \(\mu\)M in \(\mathrm{Ca}^{2+}\) signaling~\cite{Faas2011}) compared to the threshold ligand concentration (e.g., approximately \(0.5\,\mathrm{nM}\) for cAMP during threshold sensing in \emph{Dictyostelium}~\cite{Haastert2007}), implying that the associated relative fluctuations are small in intracellular components, scaling inversely with the square root of the molecule number. Therefore, by focusing on the dominant source of sensing error, namely receptor-level noise, we approximate the total sensing error in our analysis. Second, our goal is to estimate a lower bound on sensing error in \(\mathrm{Ca}^{2+}\) signaling, and including intrinsic noise is expected to increase this bound.}

The coupled nonlinear rate equations (Eq.~1 to Eq.~5) is solved analytically by linearizing them around the steady state, as detailed in Appendix \ref{appendA}.  {This procedure assumes that concentration fluctuations are small compared to their corresponding steady-state values, allowing for a approximation: \( \bar{C}_{\text{Ca}} \, \delta C_K + \bar{C}_K \, \delta C_{\text{Ca}} \gg \delta C_{\text{Ca}} \, \delta C_K \), which ensures that cross term dominate over quadratic fluctuation term. This can be equivalently expressed as \( \frac{\delta C_{\text{Ca}}}{\bar{C}_{\text{Ca}}} + \frac{\delta C_K}{\bar{C}_K} \ll 1 \). Here, \( \bar{C}_{\text{Ca}} \) and \( \bar{C}_K \) denote the steady-state concentrations of calcium and calmodulin, respectively, while \( \delta C_{\text{Ca}} \) and \( \delta C_K \) represent their corresponding fluctuations around the steady state. The fluctuation in the phosphorylation level at measurement time \( T \) is given by \( \delta C_{Pr}(T) = \int_{0}^{T} f(T - t') \, \delta p(t') \, dt' \), as derived in Appendix \ref{appendB}. The function \( f(\Delta t = T - t') \) is the weighting (response or transfer) function that determines how the processing network at time \( T \) weights the receptor-level signal at an earlier time \( t' \) (for explicit form of this function, see Eq.~\eqref{weigting_function}). The effective integration time \( T_{\mathrm{eff}} \) is determined by the temporal range over which \( f(\Delta t) \) remains nonzero~\cite{Govern2012}. The measurement time \( T \) (or observation time in the sense of \cite{Govern2012}) serves as an upper bound on \( T_{\mathrm{eff}} \). We define the timescales of cytoplasmic reactions as the inverses of the corresponding backward rate constants, such as \( 1/k_{dp} \), \( 1/k_{d1} \), and \( 1/k_{l} \)~\cite{Biswal2024}. These time scales combine together to produce a time scale that determines how long output retains memory.}

{Using the exponentially decaying autocorrelation function of \( p(t) \), given by \( G(t', t'') = \langle p(t') \, p(t'') \rangle = \bar{p}^2 + \bar{p} \, (1-\bar{p}) \, e^{-\frac{|t''-t'|}{(1-\bar{p}) \tau_b}} \)~\cite{Berg1977}}, the RMS error in estimating the mean ligand concentration \( \bar{c} \) for a single receptor over a long measurement time \( T \), as derived in Appendix \ref{appendC}, is given by
\begin{equation}
\begin{aligned}
&\frac{\Delta c_{rms}}{\bar{c}} = \, \bigg[ \frac{ d(a^2-b) \tau_c}{\bar{p} \, (1 - \bar{p})} \, \times \, \frac{4 (a + d) (1+ (a + d) \, \tau_c) + a \, [(a + 2d)^2-b] \, \tau_c^2}{a  \, [(a + 2d)^2-b] \, (1 + d \tau_c) \, [4 + 4a \tau_c + (a^2 - b) \tau_c^2]} \bigg]^{\frac{1}{2}},
\end{aligned}
\label{simplified}
\end{equation}
which is valid under the linearization condition \( \delta C_{\text{Ca}} / \bar{C}_{\text{Ca}} + \delta C_K / \bar{C}_K \ll 1 \) as discussed above. This condition translates to $\frac{I_{\text{Ca}} \, k_{k1} \, \bar{p} \, (1-\bar{p})}{2 \, I_{\text{Ca}} \, k_{k1} \, \bar{p} + k_{d1} \, k_{l}} \, \frac{\Delta c_{\text{rms}}}{\bar{c}} \ll 1 $ (see Appendix \ref{appendH} for details). Various parameters in above equations are defined as follows:
\begin{align}
    a &= k_{d1} + \Bar{C}_{Ca} k_{k1} + \Bar{C}_{K} k_{k1} + k_{l} \\ \nonumber
    b &= (k_{d1} + \Bar{C}_{Ca} k_{k1} + \Bar{C}_{K} k_{k1} + k_{l})^2  \\
    & \quad -4 (k_{d1} k_{l} + \Bar{C}_{Ca} k_{k1} k_{l}) \\
    d &= k_{dp}
\end{align} 
\begin{subequations}
\begin{align}
    \bar{C}_{\text{Ca}} &= \frac{I_{\text{Ca}} \, \bar{p}}{k_l} \\
    \bar{C}_{\text{K}} &= \frac{C_{K0} \, k_{d1} \, k_l}{k_{d1} \, k_l + I_{\text{Ca}} \, k_{k1} \, \bar{p}},
\end{align}
\end{subequations}
{where, \( \bar{p} =\frac{k_+\bar{c}}{k_+\bar{c} + k_-} \) represents the average receptor occupation probability, \( \tau_c =(1 - \bar{p}) \, \tau_b = \frac{1}{k_+\bar{c} + k_-} \) is the correlation time, reflecting the duration over which the state of a receptor retains a significant correlation with its past state, and \( \tau_b =k_-^{-1} \) goes as the average time for which the receptor is bound.}

 {Eq.~\eqref{simplified} describes the concentration sensing error in the limit of large measurement times $T$. It shows that the error approaches zero in the absence of dephosphorylation, i.e., when \( k_{dp} = 0 \), and the phosphorylated output retains information about the input signal over an extended period, corresponding to a large effective integration time. This behavior is consistent with Refs.~\cite{Berg1977, Bialek2005, Endres2008, Endres2009Concentration, Bialek2008, Skoge2011, Govern2012, Wang2007, Mora2010, Kaizu2014, Berezhkovskii2013, Irina2003, Haastert2007, Ipina2016, Wasnik2022}. In these works, a long integration time enables the system to average over many statistically independent receptor samples, thereby suppressing input fluctuations and driving the sensing error toward zero.}

 {Conversely, when \( k_{dp} \neq 0 \), the effective integration time becomes finite, and the sensing error no longer vanishes, even in the limit of long measurement time. Instead, the error saturates to a nonzero value, as also demonstrated by our stochastic simulations shown in Fig.~\ref{singlecalciumErrorVsTime} (see Appendix \ref{appendK} for simulation methodology). This saturation arises because the system's memory is constrained by the intrinsic timescales of the signaling network. Specifically, the output can only integrate the input signal over a finite period determined by the cytoplasmic deactivation rates, as discussed previously. This saturation behavior is consistent with the findings of Ref.~\cite{GovernNov2014}, where a similar effect was observed in the context of prokaryotic \textit{E.~coli} signaling networks.}
  
\begin{figure}[htpb]
    \centering
    \begin{minipage}[b]{0.30\textwidth}
        \centering
      \raggedright\textbf{(a)}  \includegraphics[width=\textwidth]{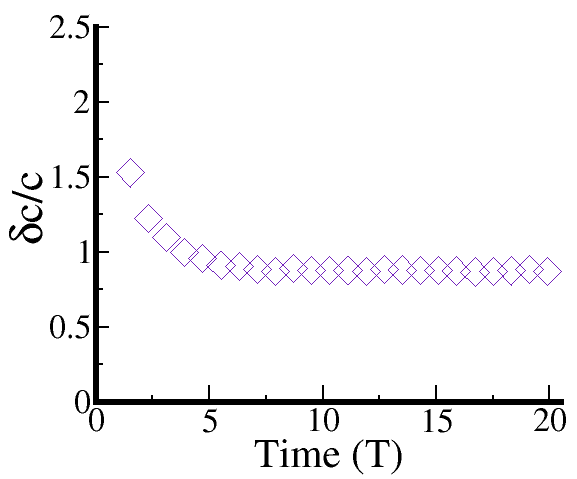}
        \label{fig:image-1a}
    \end{minipage}
    \begin{minipage}[b]{0.30\textwidth}
        \centering
       \raggedright\textbf{(b)}  \includegraphics[width=\textwidth]{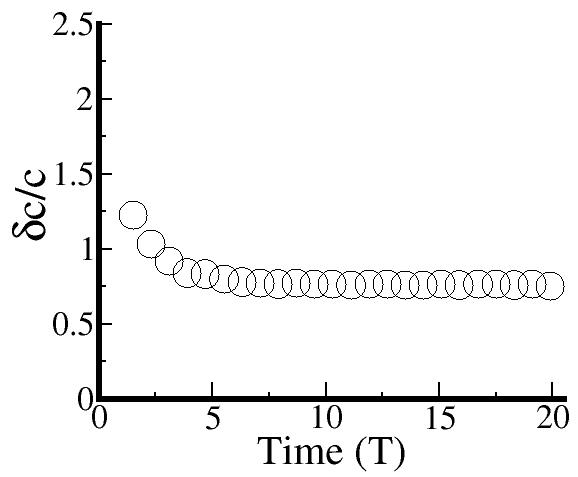}
        \label{fig:image-1b}
    \end{minipage}
    \begin{minipage}[b]{0.30\textwidth}
        \centering
      \raggedright\textbf{(c)}   \includegraphics[width=\textwidth]{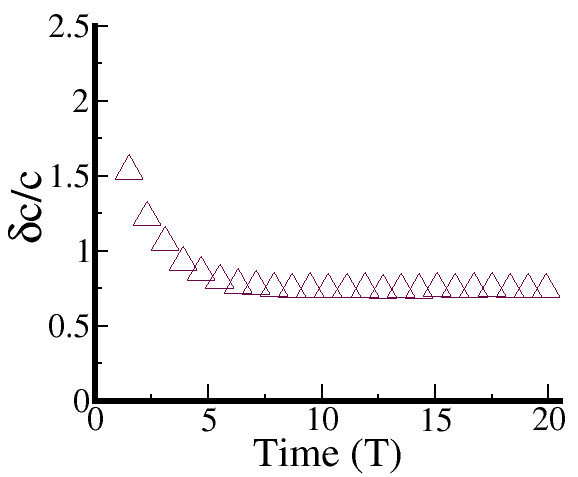}
        \label{fig:image-1c}
    \end{minipage}
    \begin{minipage}[b]{0.30\textwidth}
        \centering
       \raggedright\textbf{(d)}  \includegraphics[width=\textwidth]{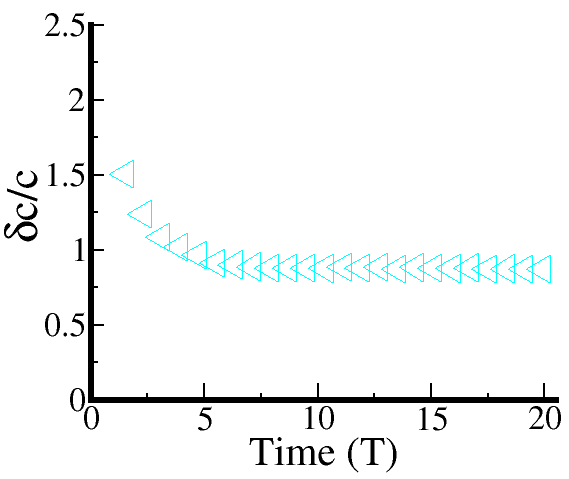}
        \label{fig:image-1d}
    \end{minipage}
    \begin{minipage}[b]{0.30\textwidth}
        \centering
       \raggedright\textbf{(e)}  \includegraphics[width=\textwidth]{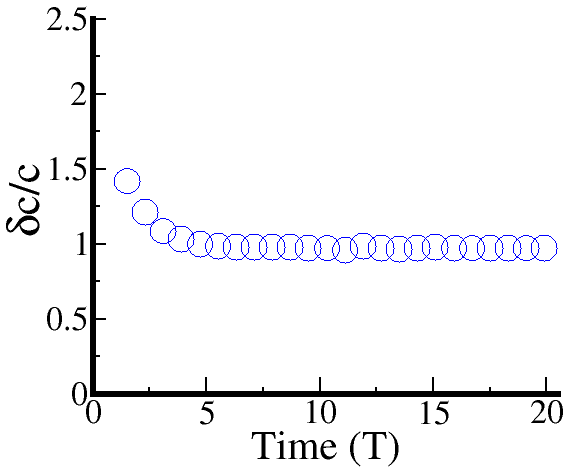}
        \label{fig:image-1e}
    \end{minipage}
    \begin{minipage}[b]{0.30\textwidth}
        \centering
        \raggedright\textbf{(f)} \includegraphics[width=\textwidth]{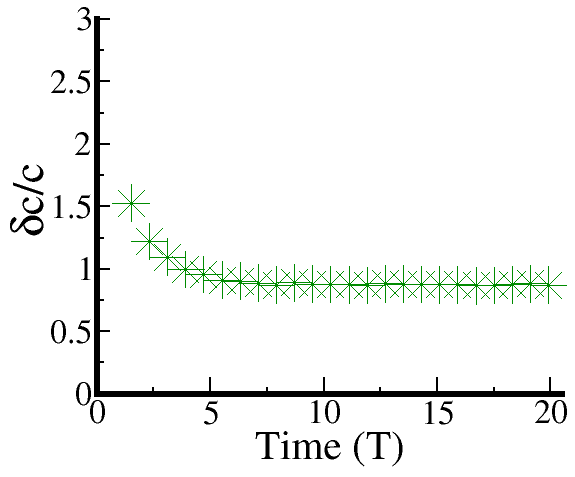}
        \label{fig:image-1f}
    \end{minipage}
    \begin{minipage}[b]{0.30\textwidth}
        \centering
      \raggedright\textbf{(g)}   \includegraphics[width=\textwidth]{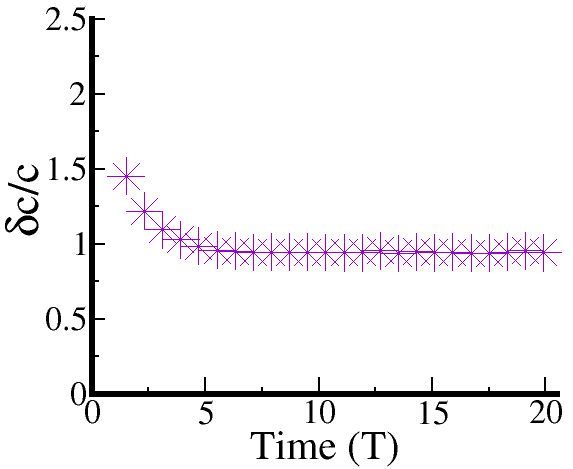}
        \label{fig:image-1g}
    \end{minipage}
    \begin{minipage}[b]{0.30\textwidth}
        \centering
      \raggedright\textbf{(h)}   \includegraphics[width=\textwidth]{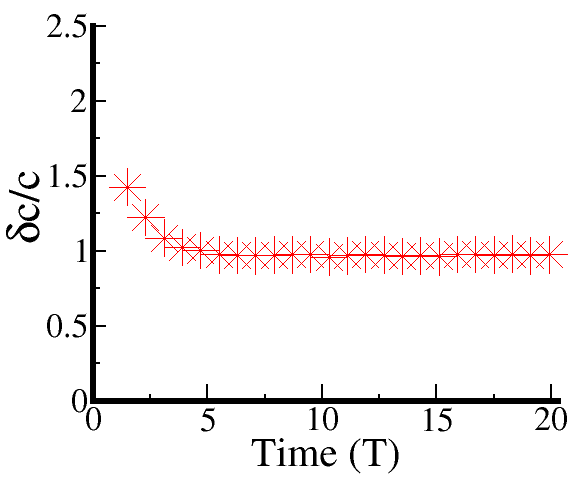}
        \label{fig:image-1h}
    \end{minipage}
  \caption{Error ($\delta c/c$) vs. measurement time ($T$) plot for cellular signaling. This plot shows how error behaves over time, demonstrating that the error saturates at a steady-state value as time progresses. The figure includes various scenarios illustrating the effect of different cytoplasmic rate constants and ligand attachment/detachment rates on the error. Units for all rate constants: $k_{k1}$ and $k_+$ are in $M^{-1}s^{-1}$; $k_-$,  $k_{d1}$, $k_p$, $k_{dp}$, and $k_l$ are in $s^{-1}$; $I_{Ca}$ is in $Ms^{-1}$. In (a), all cytoplasmic rate constants ($k_{k1}$, $k_{d1}$, $k_p$, $k_{dp}$, $k_l$), as well as ligand attachment and detachment rates, are set to 1 (units specified as above). In (b), all cytoplasmic rate constants and detachment rates are 1, but $k_+$ is $2M^{-1}s^{-1}$. In (c), $k_-$ is set to $3s^{-1}$ while all other parameters remain at 1. The effects of varying specific rate constants are shown as follows: $k_{k1}$ is $2M^{-1}s^{-1}$ in (d); $k_{d1}$ is $4s^{-1}$ in (e); $k_p$ is $4s^{-1}$ in (f); $k_{dp}$ is $2s^{-1}$ in (g); and $k_l$ is $3s^{-1}$ in (h), with all other parameters fixed at 1. Throughout these scenarios, the calcium influx rate ($I_{\text{Ca}}$) is maintained at $10^{-2}Ms^{-1}$, and the initial kinase concentration is $10^{-4}M$. The plots reveal that the steady-state error remains nonzero across different changes in cytoplasmic rate constants and ligand binding and unbinding rates.}
 \label{singlecalciumErrorVsTime}
\end{figure}

 {At large measurement times $T$, when the timescales of cytoplasmic reactions (defined as the inverses of the backward rate constants, such as \( 1/k_{dp} \), \( 1/k_{d1} \), and \( 1/k_{l} \) as discussed before) is much longer than the receptor correlation time ($\tau_c$),  {and dephosphorylation is the rate-limiting step (i.e., \( 1/k_{\text{dp}} \gg 1/k_{\ell}, 1/k_{\text{d1}} \))}, the sensing error [Eq.~\eqref{simplified}] simplifies to $\frac{\Delta c_{\text{rms}}}{\bar{c}} \approx \sqrt{k_{\text{dp}} \, \bigg( \frac{1}{k_-} + \frac{1}{k_+ \,  \bar{c}} \bigg)} = \sqrt{\frac{1}{T_{eff}} \left( \frac{1}{k_-} + \frac{1}{k_+ \, \bar{c}} \right)}$ (see Appendix \ref{appendI} for derivation and the validity of linearization approximation). This expression is analogous to the classical Berg–Purcell limit ~\cite{Berg1977} for a single receptor, given by \( \frac{\Delta c_{\text{rms}}}{\bar{c}} \approx \sqrt{\frac{2 \tau_b}{T \bar{p}}} = \sqrt{\frac{2}{T} \left( \frac{1}{k_-} + \frac{1}{k_+ \bar{c}} \right)} \) ~\cite{Kaizu2014}. In this context, the measurement time \( T \) in the Berg–Purcell framework is effectively replaced by the network’s memory timescale, i.e., the effective integration time \( T_{\text{eff}} = 1/k_{\text{dp}} \), which is determined by how rapidly dephosphorylation of the output erases memory of the input signal.}

 {On the other hand, when the timescales of cytoplasmic reactions are much shorter than $\tau_c$, Eq.~\eqref{simplified} simplifies to $\Delta c_{\text{rms}} / \bar{c} \approx \left( \bar{p}(1 - \bar{p}) \right)^{-1/2} = (k_+ \bar{c} + k_-)/\sqrt{k_+ \bar{c} \cdot k_-}$ (see Appendix \ref{appendI} for derivation and the validity of linearization approximation). This expression highlights that, unlike the Berg-Purcell prediction, the error does not necessarily decrease with increasing ligand binding and unbinding rates. Instead, the error may increase, particularly when the attachment rate substantially exceeds the detachment rate or vice versa. In this regime, the memory of intracellular components, characterized by the timescales $1/k_{dp}$, $1/k_{d1}$, and $1/k_{l}$, is too short to retain information about past receptor activity, rendering the effective integration time $T_{\text{eff}}$ negligible. Consequently, the system estimates the concentration from instantaneous fluctuations in receptor occupancy $p(t)$, without any temporal integration. This corresponds to the “instantaneous error,” i.e., the sensing error associated with a single measurement made by a single receptor~\cite{GovernNov2014}. The error attains a minimum when $k_+ \bar{c} = k_-$ (i.e., $\bar{p} = 1/2$), yielding $\Delta c_{\text{rms}} / \bar{c} = 2$. Therefore, in the absence of time integration, there exists a lower bound on the sensing error, given by $\Delta c_{\text{rms}} / \bar{c} \ge 2/\sqrt{N}$ for $N$ independent receptors. This bound is consistent with previous results for steady-state concentration sensing~\cite{GovernDec2014, tenWolde2016}. This limiting behavior also corresponds to the asymptotic regime described by~\cite{Malaguti2021}, who studied the accuracy of sensing time-varying ligand concentrations that have a correlation time $\tau_L$. As shown in their Figure~3(d), increasing the receptor correlation time $\tau_c$ of ligand binding leads to more strongly correlated receptor samples. To maintain effective time integration, i.e., to accumulate a sufficient number of statistically independent samples, the receptor integration time $\tau_r$ (set by the relaxation time of the push-pull network, analogous to $T_{\text{eff}}$ in our framework) initially increases with $\tau_c$, as the number of independent receptor samples scales as $\tau_r/\tau_c$. However, once $\tau_r$ approaches the signal correlation time $\tau_L$, further increase leads to signal distortion: the system begins to average over variations in the signal itself, causing the output to lag and no longer reflect the instantaneous concentration. To avoid such distortion, the optimal integration time $\tau_r$ then decreases with increasing $\tau_c$ and eventually vanishes—leading to a strategy where the system ceases time integration altogether and becomes an instantaneous responder. Thus, the asymptotic regime shown in Fig.~3(d) of~\cite{Malaguti2021} corresponds precisely to the limit we analyze, where $\tau_r$ (or cytoplasmic timescales in our terminology) becomes much shorter than $\tau_c$, and the system behaves as an instantaneous responder. Although~\cite{Malaguti2021} considers time-varying ligand concentrations, whereas our study focuses on static ligand concentrations, this distinction becomes irrelevant in the limit where the optimal integration time $\tau_r \to 0$. In this regime, the system responds on a timescale much shorter than the signal's correlation time $\tau_L$, effectively reducing the sensing problem to that of instantaneous concentration estimation.}

The lower bound for the RMS error in concentration estimation is derived  by identifying the minima of the numerator and the maxima of the denominator of Eq.~\eqref{simplified} through term reduction, and is given by (see Appendix \ref{appendD} for details)
\begin{align}
  \frac{\Delta c_{rms}}{\bar{c}} &> \left[ \frac{I_{Ca} k_{k1} k_{dp}}{(k_+ \, \bar{c} + k_- ) + k_{dp}} \right]^{\frac{1}{2}} \times \frac{1}{k_{d1} + \frac{I_{\text{Ca}} k_{k1}}{k_l} + C_{k0} k_{k1} + k_{l} + 2(k_+ \, \bar{c} + k_-)}.\label{lower bound}
\end{align}
This expression represents the lower bound of the error in sensing ligand concentration through calcium signaling once the cell has reached a steady state, under the assumptions of the linearization approximation used to develop Eq.~\ref{simplified}.

\begin{figure}[htbp]
    \centering
    \begin{minipage}[b]{0.30\textwidth}
        \centering
      \raggedright\textbf{(a)}  \includegraphics[width=\textwidth]{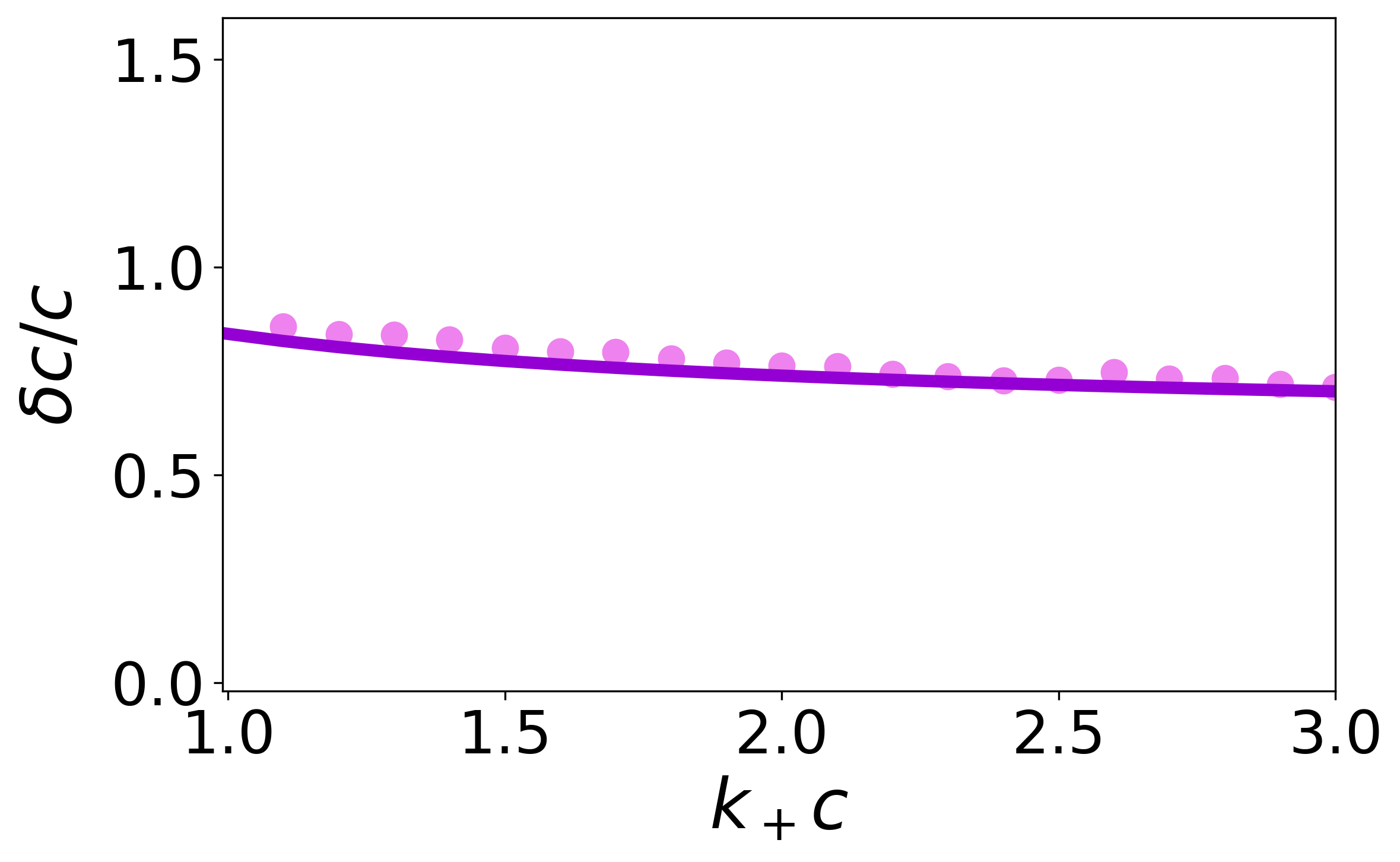}
        \label{fig:image-3a}
    \end{minipage}
    \begin{minipage}[b]{0.30\textwidth}
        \centering
       \raggedright\textbf{(b)}  \includegraphics[width=\textwidth]{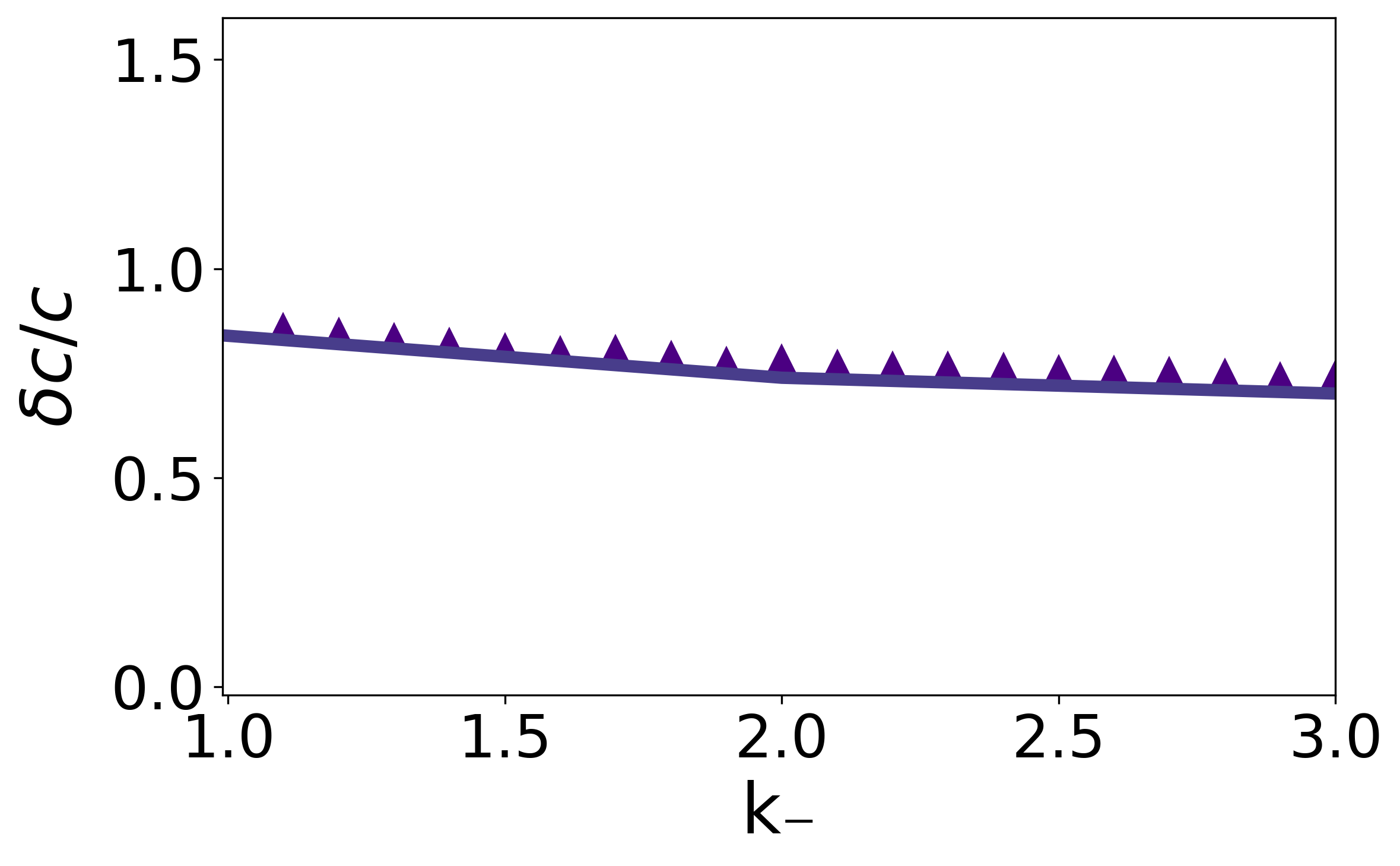}
        \label{fig:image-3b}
    \end{minipage}
    \begin{minipage}[b]{0.30\textwidth}
        \centering
       \raggedright\textbf{(c)}  \includegraphics[width=\textwidth]{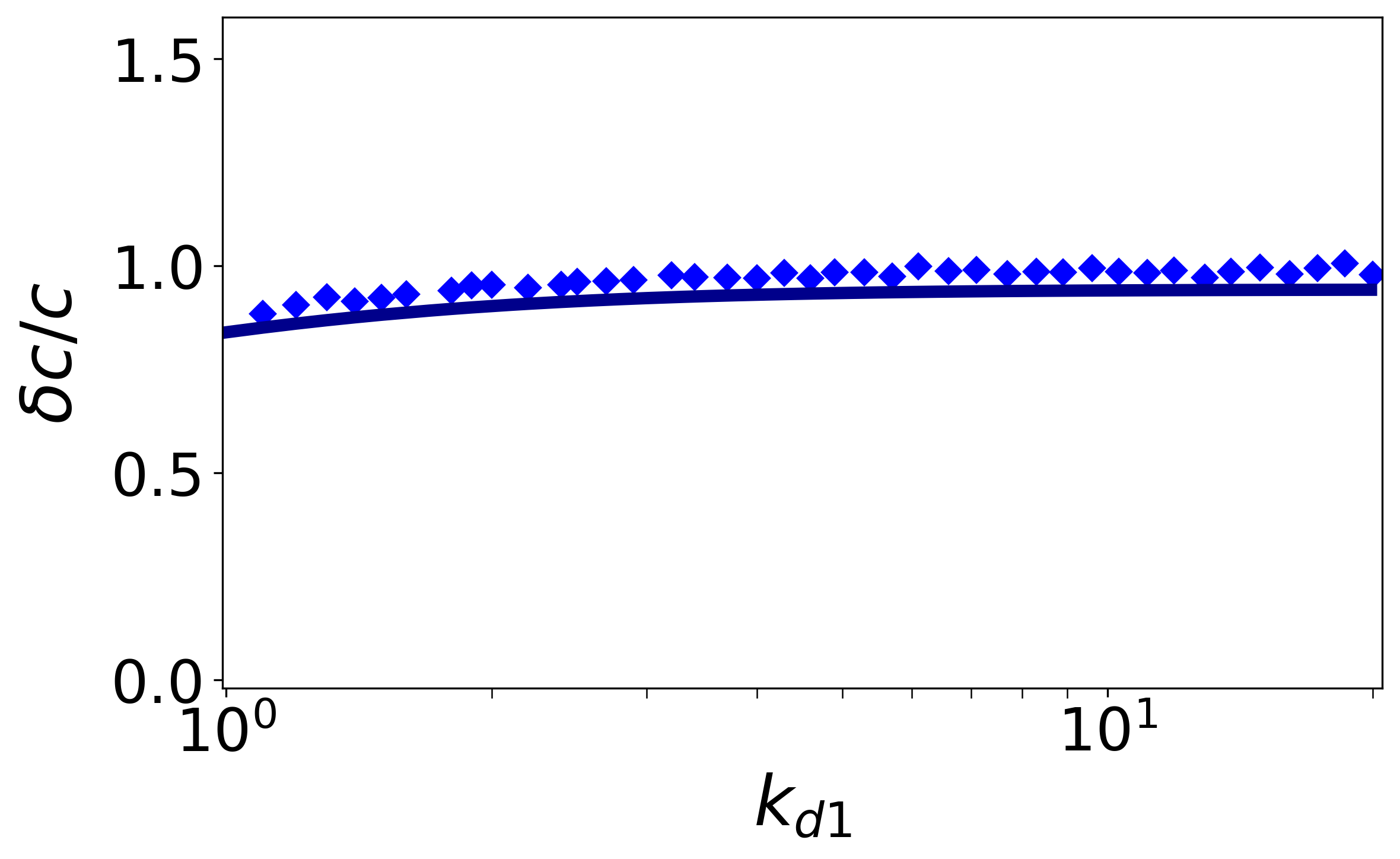}
        \label{fig:image-3c}
    \end{minipage}
    \begin{minipage}[b]{0.30\textwidth}
        \centering
       \raggedright\textbf{(d)}  \includegraphics[width=\textwidth]{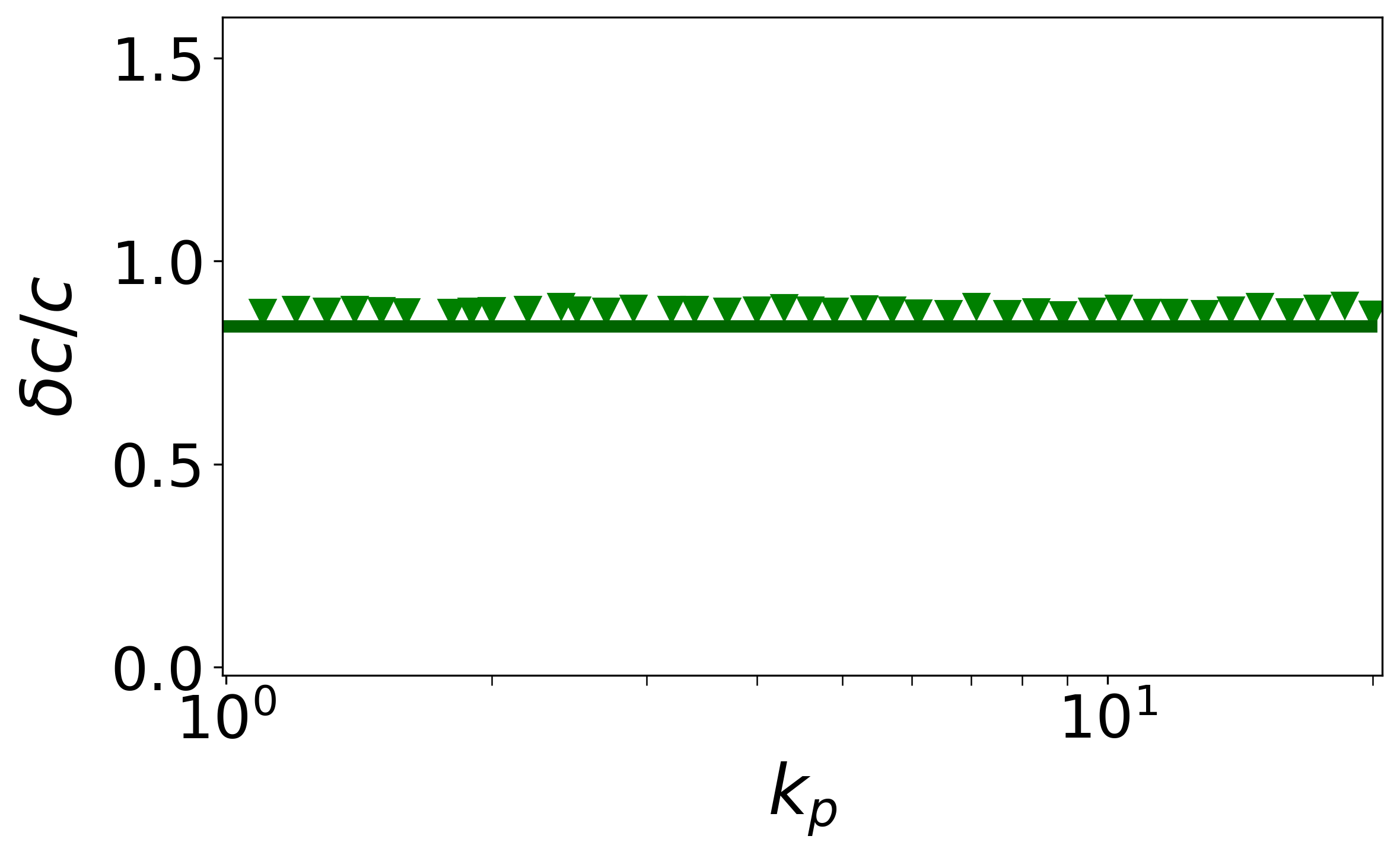}
        \label{fig:image-3d}
    \end{minipage}
    \begin{minipage}[b]{0.30\textwidth}
        \centering
        \raggedright\textbf{(e)} \includegraphics[width=\textwidth]{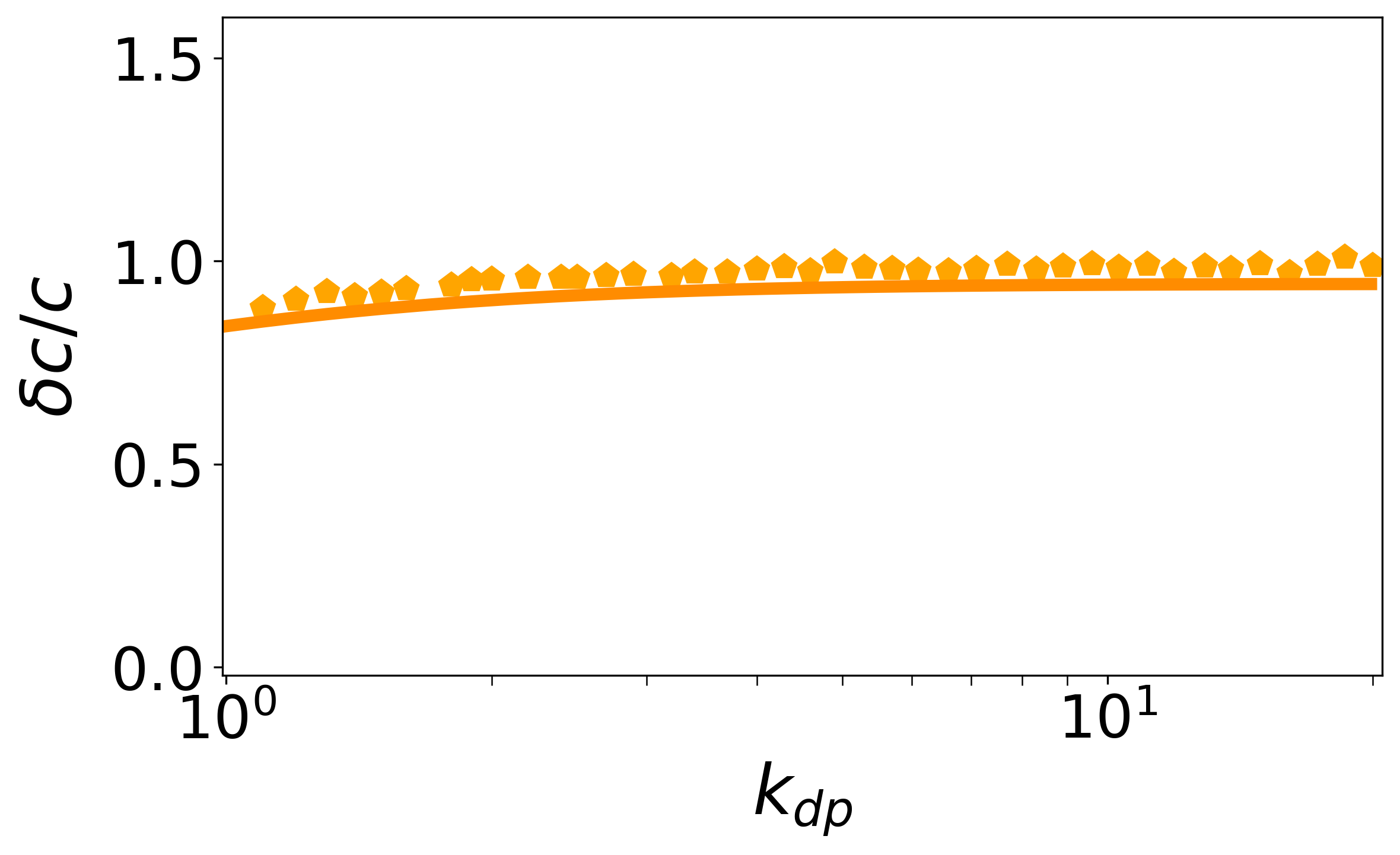}
        \label{fig:image-3e}
    \end{minipage}
    \begin{minipage}[b]{0.30\textwidth}
        \centering
      \raggedright\textbf{(f)}   \includegraphics[width=\textwidth]{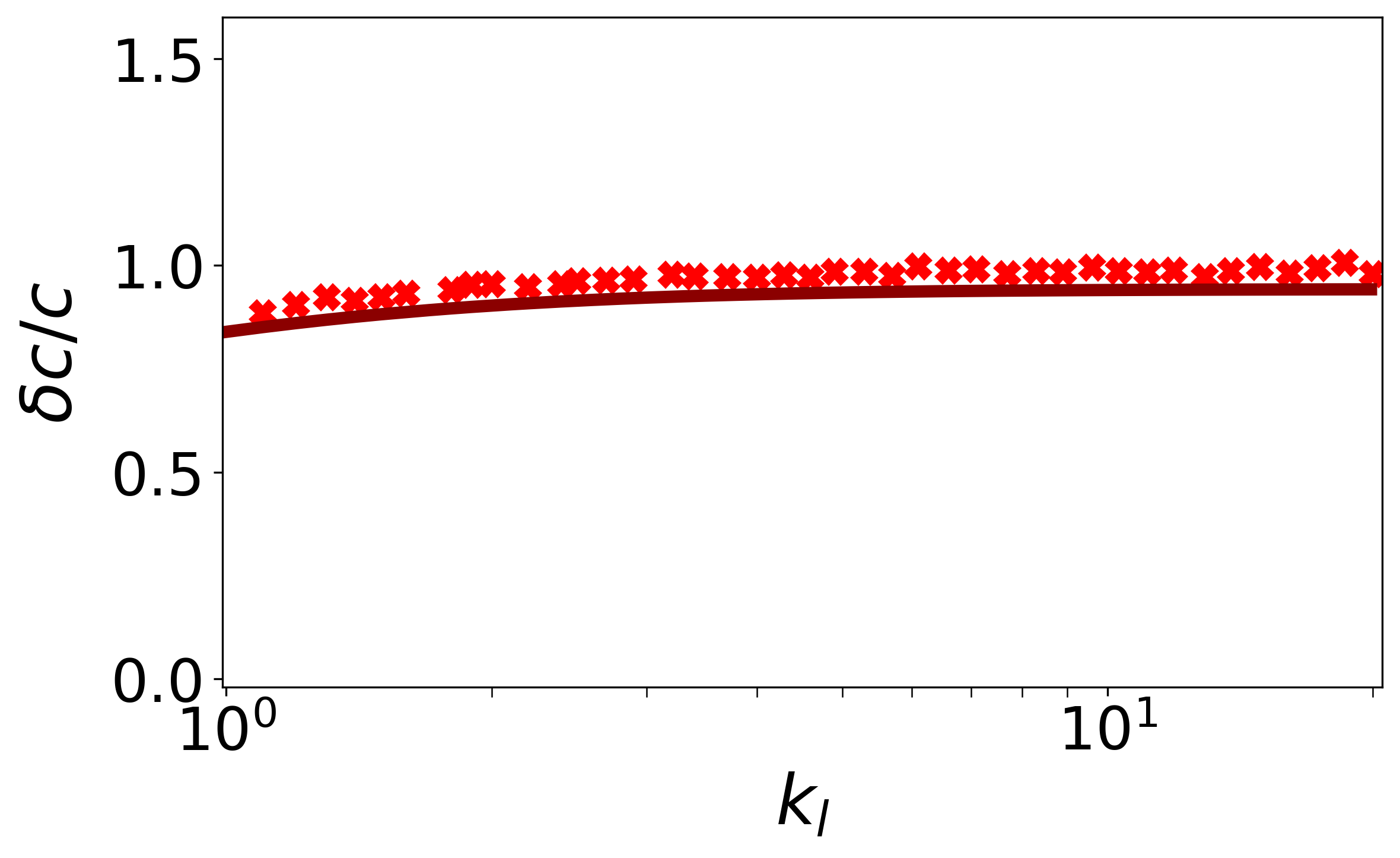}
        \label{fig:image-3f}
    \end{minipage}
 \caption{Plots of error ($\delta c / c$) vs. various rate constants: (a) ligand attachment ($k_+ c$), (b) ligand detachment ($k_-$), (c) K-Ca dissociation ($k_{d1}$), (d) phosphorylation ($k_p$), (e) dephosphorylation ($k_{dp}$), and (f) calcium loss ($k_l$). All other cytoplasmic rate constants ($k_{k1}$, $k_{d1}$, $k_p$, $k_{dp}$, $k_l$) are fixed at 1 (with \(k_{k1}\) in units of \(M^{-1}s^{-1}\) and the others in \(s^{-1}\) ), except for the varying one in each plot. $\bar{c}$, $k_+$ and $k_-$ are also set to $1M$, $1M^{-1}s^{-1}$ and $1s^{-1}$ respectively. The solid line in each plot represents the analytic result [Eq.~\eqref{simplified}], while the data points show the simulation result. The calcium influx rate ($I_{Ca}$) is set to  $10^{-2}Ms^{-1}$, measurement time ($T$) is set to $10s$, and initial kinase concentration is set to $10^{-4}M$.  {Panels (c)–(f) use a logarithmic horizontal axis to illustrate the sensitivity of the error to variations in parameter values across several orders of magnitude. In contrast, panels (a) and (b) are plotted with a linear horizontal axis, as the explored parameter range is narrow (from 1 to 3). Increasing \(k_-\) significantly beyond this range, while keeping \(k_+ c = 1\), would reduce the steady-state receptor occupancy \( \bar{p} \), thereby diminishing the signal (mean concentrations). This leads to a breakdown of the linearization condition stated below Eq.~\eqref{simplified}, which requires sufficiently large signal amplitudes relative to fluctuations. A broader range of \( k_+ c \) and \( k_- \) values is explored in Fig.~1 and Fig.~2 of the SI, respectively. However, those plots correspond to a different set of parameters than those used here.}}
    \label{fig:errorVsrateconstplots}
\end{figure}

 {Fig.~\ref{fig:errorVsrateconstplots} illustrates the agreement between the simulations and the analytical result (Eq.~\ref{simplified}) and how various rate constants influence the error in sensing the extracellular ligand concentration. Figure~\ref{fig:errorVsrateconstplots}(a) shows that the sensing error initially decreases with increasing attachment rate \( k_+ \bar{c} \), and then saturates at a finite value. This behavior can be understood by viewing the signaling network as a sampling device ~\cite{GovernNov2014}, and therefore rewriting Eq.~\eqref{simplified} in the form \( \left( \Delta c_{\mathrm{rms}} / \bar{c} \right)^2 = \left[ \bar{p}(1 - \bar{p}) \left(1 + T_{\mathrm{eff}}/\tau_c \right) \right]^{-1} \). In this interpretation, \(1/\bar{p}(1 - \bar{p})\) captures the instantaneous error, as discussed previously, i.e., the sensing error associated with a single receptor measurement in the absence of temporal integration. The factor \(1 + T_{\mathrm{eff}}/\tau_c\) quantifies the number of effectively independent receptor samples acquired over the integration window \(T_{\mathrm{eff}}\). Here, \(T_{\mathrm{eff}}\) is governed by cytoplasmic relaxation processes and is set by \(1/k_{dp}\), \(1/k_{d1}\), and \(1/k_{l}\) as discussed before. As \(k_+ \bar{c}\) increases, receptor switching becomes faster, decreasing the receptor correlation time \( \tau_c = 1 / (k_+ \bar{c} + k_-) \), and thereby increasing the number of independent samples within the integration window, which leads to improved sensing precision. However, in the limit where \(k_+ \bar{c} \gg k_-\), the receptor becomes almost fully occupied (\(\bar{p} \to 1\)), so that \(1 - \bar{p} \to 0\), and \(T_{\mathrm{eff}} / \tau_c \gg 1\) (as \(\tau_c\) becomes very small). In this regime, the error simplifies to \( \left( \Delta c_{\mathrm{rms}} / \bar{c} \right)^2 \approx \left[(1 - \bar{p}) \left( T_{\mathrm{eff}}/\tau_c \right) \right]^{-1} \approx \left( k_- T_{\mathrm{eff}} \right)^{-1} \), showing that the sensing error is no longer limited by the binding rate \(k_+ \bar{c}\), but instead by the unbinding rate \(k_-\) and the duration of integration. As a result, the error saturates for large \(k_+ \bar{c}\), consistent with the asymptotic behavior observed in the figure. A similar explanation applies to Figure~\ref{fig:errorVsrateconstplots}(b), where the RMS error initially decreases with increasing unbinding rate \(k_-\), and then reaches a plateau once the error becomes limited by the binding dynamics and the effective integration time.}

The independence of the sensing error from the phosphorylation rate \( k_p \) (~\ref{fig:errorVsrateconstplots}d) arises from the fact that the expression for the RMS error in Eq.~(6) does not contain \( k_p \). This can be directly understood from the rate equation governing the phosphorylated output concentration, \( C_{\text{Pr}} \), given by Eq.~(5): \( \frac{dC_{\text{Pr}}(t)}{dt} = k_p C_{K\text{-}Ca}(t) - k_{dp} C_{\text{Pr}}(t) \), whose solution at measurement time \( T \) is \( C_{\text{Pr}}(T) = k_p \int_0^T e^{-k_{dp}(T - t')} C_{K\text{-}Ca}(t') \, dt' \), indicating a linear dependence on \( k_p \). As both the mean and standard deviation of \( C_{\text{Pr}} \) scale linearly with \( k_p \), the RMS error of output $\sigma_{{C}_{\text{Pr}}}/ \bar{C}_{\text{Pr}}$, and hence the ligand concentration sensing error \( \Delta c_{\text{rms}} / \bar{c} \), becomes independent of \( k_p \).  {Although increasing \( k_p \) amplifies the output signal by increasing the number of phosphorylated molecules, it also proportionally increases the fluctuations in the output. As a result, the signal-to-noise ratio (SNR) ratio remains unchanged. This observed independence of the sensing error from \( k_p \) relies on the assumption of large copy numbers of cytoplasmic components, where intrinsic noise can be neglected. However, when intrinsic fluctuations become significant, as in genetic regulatory networks with low copy numbers, the sensing error can depend nontrivially on \( k_p \)~\cite{deRonde2010, Tostevin2009}. In contrast, the expression \( C_{\text{Pr}}(T) = k_p \int_0^T e^{-k_{dp}(T - t')} C_{K\text{-}Ca}(t') \, dt' \) shows that the dephosphorylation rate \( k_{dp} \) enters nonlinearly, affecting both the mean and the standard deviation of \( C_{\text{Pr}} \) nonlinearly. As a result, its contribution does not cancel in the RMS error. This distinction can be further understood using the weighting function \( f(\Delta t) \), which defines the effective integration time \( T_{\mathrm{eff}} \). As discussed earlier, \( T_{\mathrm{eff}} \) is determined by the temporal range over which \( f(\Delta t) \) remains nonzero, and Eq.~(C.10) in the SI shows that \( f(\Delta t) \) depends linearly on \( k_p \). However, since the relative error depends on the normalized form of \( f(\Delta t) \)~\cite{Govern2012}, any linear dependence on \( k_p \) is eliminated, meaning \( k_p \) does not influence the effective temporal window. Conversely, \( f(\Delta t) \) depends nonlinearly on \( k_{dp} \), and thus the normalized weighting function retains its dependence on \( k_{dp} \), directly influencing the effective integration time \( T_{\mathrm{eff}} \). Specifically, as shown in Eq.~\eqref{weigting_function}, \( f(\Delta t) \) decays exponentially with increasing \( k_{dp} \), and this decay dominates even after normalization, since other linear dependencies are suppressed in the normalized form. As a result, increasing \( k_{dp} \) reduces \( T_{\mathrm{eff}} \), thereby increasing the RMS error initially. Once \( k_{dp} \) becomes significantly larger than the other cytoplasmic rates, the slower processes begin to dominate the system's dynamics. As a result, the temporal profile of  \( f(\Delta t) \), and consequently the \( T_{\mathrm{eff}} \), becomes limited by these slower rates, leading to the saturation of the RMS error, as shown in Fig.~\ref{fig:errorVsrateconstplots}(e). A similar interpretation holds for the calcium unbinding rate from calmodulin \( k_{d1} \) (Fig.~\ref{fig:errorVsrateconstplots}c) and calcium leaving rate from the cytoplasm \( k_l \) (Fig.~\ref{fig:errorVsrateconstplots}f), where the RMS error saturates once these rates exceed other cytoplasmic reaction rates. This saturation behavior is also captured in the analytic expressions provided in Eq.~\eqref{largerkd1} and~\eqref{largekl}. For instance, Eq.~\eqref{largerkd1} demonstrates that when \( k_{d1} \) becomes much larger than the other cytoplasmic rates, the RMS error ceases to be limited by \( k_{d1} \) and is instead determined by other rates.}
 
{The RMS error in sensing ligand concentration can be derived under the assumption that one cytoplasmic rate constant dominates the others, subject to the constraint: $\bar{C}_{\text{Ca}} \, \delta C_K + \bar{C}_K \, \delta C_{\text{Ca}} \gg \delta C_{\text{Ca}} \, \delta C_K$.}

{When \( k_{k1} \) is significantly larger than the other cytoplasmic rate constants, the RMS error, derived in Appendix \ref{appendE}, is given by:}
\begin{equation}
\frac{\Delta c_{rms}}{\bar{c}} = \left[ \frac{k_{dp} k_{l} \tau_c}{\bar{p} \, (1 - \bar{p})} \, \frac{(1 + k_{l} \tau_c + k_{dp} \tau_c)}{(k_{dp} + k_{l})(1 + k_{dp} \tau_c)(1 + k_{l} \tau_c)} \right]^{\frac{1}{2}}.
\label{largerkk1}
\end{equation}

{If \( k_{d1} \) is much larger than the other rate constants, the RMS error follows the same expression as Eq.~\eqref{largerkk1} (see Appendix \ref{appendE} for details):}
\begin{equation}
\frac{\Delta c_{rms}}{\bar{c}} = \left[ \frac{k_{dp} k_{l} \tau_c}{\bar{p} \, (1 - \bar{p})} \, \frac{(1 + k_{l} \tau_c + k_{dp} \tau_c)}{(k_{dp} + k_{l})(1 + k_{dp} \tau_c)(1 + k_{l} \tau_c)} \right]^{\frac{1}{2}}.
\label{largerkd1}
\end{equation}
{When \( k_{l} \) is the largest rate constant, the RMS error is given by (see Appendix \ref{appendE} for details):}
\begin{equation}
\frac{\Delta c_{rms}}{\bar{c}} = \left[ \frac{k_{dp} k_{d1} \tau_c}{\bar{p} \, (1 - \bar{p})} \, \frac{(1 + k_{d1} \tau_c + k_{dp} \tau_c)}{(k_{dp} + k_{d1})(1 + k_{dp} \tau_c)(1 + k_{d1} \tau_c)} \right]^{\frac{1}{2}}.
\label{largekl}
\end{equation}
When either \(k_{k1}\) or \(k_{d1}\) are large, Eq.~\eqref{largerkk1} and Eq.~\eqref{largerkd1} are equal, becoming dependent on \(k_l\) and \(k_{dp}\), while independent of \(k_{k1}\) and \(k_{d1}\). Similarly, for large \(k_l\), Eq.~\eqref{largekl} depends only on \(k_{d1}\) and \(k_{dp}\), becoming independent of \(k_l\). This independence of the sensing error from a particular cytoplasmic reaction rate, when that rate becomes significantly larger than the others, is also observed in stochastic simulations [Figs.~\ref{fig:errorVsrateconstplots}(c), \ref{fig:errorVsrateconstplots}(e), and \ref{fig:errorVsrateconstplots}(f) for $k_{d1}$, $k_{dp}$, and $k_{l}$, respectively]. This result demonstrates that concentration sensing remains robust to variations in cytoplasmic rate constants once they exceed a certain value.

%........Dicty Cell.................

%........Dicty Cell table.................
\begin{table*}
	\label{table:empirical}
	\caption{\label{table:empirical}Empirical observations for \textit{Dictyostelium} highlighting key parameters.}
	\begin{ruledtabular}
		\begin{tabular}{lll}
			Parameters & Mato et al. \cite{Mato1975} & Haastert et al. \cite{Haastert2007} \\ 
			\hline
			Threshold cAMP concentration ($c$) & 4.3 nM & 0.5 nM \\ 
			Dissociation constant ($K_D = k_-/k_+$) & $10^{-8} \, \text{M}$ & $10^{-7} \, \text{M}$ \\ 
			Total cAMP receptors  & $10^6$ & 40,000 \\ 
			Occupied receptors ($N$) at threshold sensing & 1,300 & 200 \\ 
			Occupied receptor difference between two ends & 12 & 5 \\
			Sensitivity to gradient $(\Delta c)_{\text{across cell}}/c$ & 0.9--3\% & 1--5\% \\ 
		\end{tabular}
	\end{ruledtabular}
\end{table*}

\begin{table*}
	\caption{Theoretical results for \textit{Dictyostelium}, highlighting sensing errors and gradient sensitivity.}
	\begin{ruledtabular}
		\begin{tabular}{lll}
			Sensing Error & Using Mato et al. \cite{Mato1975} & Using Haastert et al. \cite{Haastert2007} \\ 
			\hline
			$\Delta c_{\text{rms}}/\bar{c}$ for single receptor & 1.27 & 9.18 \\ 
			$\Delta c_{\text{rms}}/\bar{c}$ for $N$ receptors & 0.035 (3.5\%) & 0.689 (68.9\%) \\ 
			$(\Delta c)_{\text{across cell}}/c$, sensitivity to gradient   & 0.0702 (7.02\%) & 1.299 (129.9\%) \\ 
		\end{tabular}
	\end{ruledtabular}
	\label{table:theory}
\end{table*}

%........tables ends

 {\textit{Dictyostelium discoideum}, a eukaryotic cell, employs $Ca^{2+}$ signaling in various processes, including chemotaxis \cite{Yumura1996, Nebl1997}, mechanosensing \cite{Hashimura2022}, mechano-chemical sensing \cite{Lombardi2008}, cell differentiation \cite{Poloz2012}, and other critical cellular functions. These cells exhibit a significant chemotactic response at the threshold concentration of cyclic adenosine monophosphate (cAMP) as outlined in Table~\ref{table:empirical}. The typical calcium influx during signaling is $I_\text{Ca} = 10^{-2} \, \text{Ms}^{-1}$ \cite{Biswal2023}, while the leaving rate is $k_l = 66.67 \, \text{s}^{-1}$ \cite{clapham}. The intracellular calmodulin (CaM) concentration, $C_\text{K0}$, ranges from 47 $\mu$M to 187 $\mu$M \cite{Faas2011} (we assume \( C_{K0} = 100 \, \mu\text{M} \) for our analysis), with association and dissociation rates for calcium binding as $k_{k1} = 6.8 \times 10^6 \, \text{M}^{-1} \, \text{s}^{-1}$ and $k_{d1} = 68 \, \text{s}^{-1}$ \cite{Keller2008}. The phosphorylation rate and dephosphorylation rate is given by $k_p = 1.17 \, \text{s}^{-1}$ and $k_{dp} = 0.80 \, \text{s}^{-1}$ respectively \cite{Hong2013}. The detachment rates, $k_{-}$, for cAMP from receptors range from $0.16 \, \text{s}^{-1}$ to $1.1 \, \text{s}^{-1}$ \cite{Ueda2001}, and we assume $k_{-} = 1.1 \, \text{s}^{-1}$ for analysis. Substituting these parameters along with experimental data for from Mato et al.~\cite{Mato1975} and Haastert et al.~\cite{Haastert2007} (Table~\ref{table:empirical}) into Eq.~\eqref{simplified}, and using \( \frac{\Delta c_{\text{rms}}}{\bar{c}} \big|_N = \frac{1}{\sqrt{N}} \frac{\Delta c_{\text{rms}}}{\bar{c}} \) for \(N\) independent occupied receptors at the threshold concentration, we present our theoretical results in Table~\ref{table:theory}. Our result estimates a threshold concentration sensing accuracy of 3.5\% using data from \cite{Mato1975}, while \cite{Haastert2007} yields 68.9\%, indicating higher accuracy in the former. While the concentration sensing limits evaluated from \cite{Haastert2007} do not provide an exact concentration value, they are sufficient to distinguish whether the concentration is relatively low or high.}

 {\textit{Dictyostelium} cells are sufficiently large to simultaneously detect chemotactic signals at multiple points across their surface. By comparing these spatially distributed signals, the cells can effectively sense and respond to chemical gradients. We quantify \textit{Dictyostelium}'s sensitivity to spatial gradients using the ratio \( (\Delta c)_{\text{across cell}}/c \), following the approach in~\cite{Mato1975}. Here, \( (\Delta c)_{\text{across cell}} \) denotes the difference in extracellular cAMP concentration between the front and back halves of the cell, and \( c \) represents the background (or threshold) concentration of cAMP that the cell is capable of sensing. Successful chemotaxis requires that \textit{Dictyostelium} reliably detect this differential signal amidst stochastic fluctuations. As derived in Appendix \ref{appendJ}, the gradient sensitivity can be expressed as  
$\frac{\sigma_{\Delta c}}{c}
= \sqrt{ \frac{4N}{(N + \Delta N_{fb})(N - \Delta N_{fb})} } 
\, \cdot \, \frac{\Delta c_{\text{rms}}}{\bar{c}}$, 
where \( N \) is the total number of occupied receptors, and \( \Delta N_{fb} \) is the difference in the number of occupied receptors between the front and back halves of the cell. The term \( \Delta c_{\text{rms}} / \bar{c} \) represents the RMS error in concentration estimation by a single receptor, as given by Eq.~\eqref{simplified}. Using this formulation and the empirical values reported in Table~\ref{table:empirical}, we compute the gradient sensitivity, with the results summarized in Table~\ref{table:theory}. Our result estimates sensitivity to gradient is 7.02\% using \cite{Mato1975} and 129.9\% using \cite{Haastert2007}, implying gradient sensing feasibility only in the former. This discrepancy arises from our assumption of no cytoplasmic concentration gradient, preventing gradient measurement in our model. However, experimental evidence confirms that \textit{Dictyostelium} can sense shallow gradients~\cite{Mato1975, Haastert2007} (Table~\ref{table:empirical}). In the case of Ref.~\cite{Mato1975}, the reported threshold concentration for gradient detection is relatively high (4.3~nM), and hence does not correspond to the extreme sensitivity limit suggested by experimental measurements. As a result, ligand-receptor fluctuations are not dominant in this regime, and accurate gradient sensing is possible without accounting for cytoplasmic spatial concentration differences. By contrast, in Ref.~\cite{Haastert2007}, the threshold concentration is much lower (0.5~nM), making receptor-ligand fluctuations more significant. Under these conditions, relying solely on receptor-level measurements becomes insufficient for reliable gradient detection. Instead, intracellular spatial gradients—formed through cytoplasmic signaling—appear to play a crucial role by providing additional spatial information that compensates for the distortion of the extracellular gradient at the cell surface due to receptor noise. These observations suggest that future theoretical models should explicitly incorporate intracellular spatial heterogeneity in order to accurately describe the limits of gradient sensing and threshold detection in highly sensitive systems such as \textit{Dictyostelium}.}

%=====================discussion================================

\section{Discussion}

We evaluated how a cell senses static ligand concentrations through calcium signaling, a process ubiquitous in most eukaryotic cells. In our model, a single $Ca^{2+}$ ion binds directly to a calmodulin to activate it, and the resulting complex catalyzes the phosphorylation of target proteins, thereby simplifying the signaling network by omitting additional biochemical steps. This simplification allows us to focus on the lower bound of sensing accuracy. Including additional biochemical steps, such as the activation of calmodulin through the binding of four \( \mathrm{Ca}^{2+} \) ions followed by the activation of protein kinases, would likely introduce further sources of noise and increase the sensing error. 

Our analysis focuses on quantifying the RMS error in sensing the ligand concentration ($c$) by a single receptor, using the instantaneous level of the phosphorylated output species ($C_{\text{Pr}}$) as the readout. For $N$ independent copies of the receptor, the RMS error is scaled as $1/\sqrt{N}$. We analytically solved the coupled nonlinear rate equations  {by linearizing them under the assumption of small fluctuations around the steady-state values of intracellular component concentrations. This approach yielded an analytical expression for the RMS sensing error [Eq.~\eqref{simplified}] in the limit of long measurement times. The result shows that the sensing error saturates when the dephosphorylation rate is nonzero, as this imposes a finite effective integration time over which the output retains memory of the input signal. Stochastic simulations further support this behavior. Importantly, this saturation behavior is not unique to calcium signaling networks, which are ubiquitous in eukaryotic systems, but rather reflects a general feature of stationary signaling networks that estimate ligand concentration based on instantaneous output. In such systems, the effective integration time is limited by the internal dynamics of the network. A similar saturation effect was also observed in the context of prokaryotic \textit{E.~coli} signaling networks~\cite{GovernNov2014}.}

We derived a lower bound on the RMS error [Eq.~\eqref{lower bound}], which highlights the fundamental limit of sensing accuracy achievable through calcium signaling at steady state. We also explored cases when one cytoplasmic rate constant is dominant than other cytoplasmic rate constants. This resulted in simplified expressions [Eqs.~\eqref{largerkk1}, \eqref{largerkd1}, \eqref{largekl}] for the RMS error under specific conditions, illustrating how different rate constants impact measurement precision. These results show that concentration sensing remains robust against variations in cytoplasmic rate constants once they exceed a certain value, meaning that the sensing error becomes insensitive to a particular rate constant when it is sufficiently large compared to the others in the system. Importantly, this result is supported by both stochastic simulations and analytical calculations. The observed robustness offers a potential evolutionary advantage: if evolutionary pressures favor increased reaction rates to enhance specific cellular functions, such as the precise positional determination of the $Ca^{2+}$ entry site, the associated positional sensing error decreases with increasing \(k_{d1}\) and \(k_l\)~\cite{Wasnik2019PRL, Wasnik2019PRE}. Under such optimization, nature has the flexibility to increase these rate constants without compromising the accuracy of concentration sensing.

Our findings underscore the complexity of cellular signaling mechanisms and their implications for cellular decision-making in response to external ligand concentrations,  motivating further exploration of these intricate dynamics.

\section*{Funding}
\noindent 
Swoyam was supported by the SERB Grant No: EEQ/2021/000006.

%..........\appendix....................

\appendix

\section{Linearized Rate Equations}
\label{appendA}

The coupled nonlinear rate equations, given by Eqs. (1)–(5), can be linearized around the steady state. This approach relies on the assumption that fluctuations in the concentrations of various components are small relative to their mean values, i.e., \(\delta \mathbf{C}(t) \ll \langle \mathbf{C} \rangle\). Similarly, the occupation probability of the receptor, \( p(t) \), fluctuates around its mean value.  Thus, the steady-state values of \( p(t) \) and the concentrations can be expressed as:
\begin{align}
p(t) &= \langle p \rangle + \delta p(t) \nonumber\\
C_{\text{Ca}}(t) &= \langle C_{\text{Ca}} \rangle + \delta C_{\text{Ca}}(t) \nonumber\\
C_{\text{K}}(t) &= \langle C_{\text{K}} \rangle + \delta C_{\text{K}}(t) \nonumber\\
C_{\text{K-Ca}}(t) &= \langle C_{\text{K-Ca}} \rangle + \delta C_{\text{K-Ca}}(t) \nonumber\\
C_{\text{Pr}}(t) &= \langle C_{\text{Pr}} \rangle + \delta C_{\text{Pr}}(t).
\label{steady_state_values}
\end{align}

Substituting the above equation into Eq.~2 yields:
\begin{equation}
\begin{aligned}
\frac{d}{dt}(\langle C_{Ca} \rangle + \delta C_{Ca}) &= \bigg(I_{Ca} \, \langle p \rangle - k_l \langle C_{Ca} \rangle - k_{k1} \langle C_K \, C_{Ca} \rangle + k_{d1} \langle C_{K-Ca} \rangle \bigg) + I_{Ca} \, \delta p - k_l \, \delta C_{Ca} - k_{k1} \, \langle C_{K} \rangle  \,  \langle  C_{Ca} \rangle \\
& \quad  - k_{k1} \, \langle C_{Ca} \rangle  \, \delta C_{K} - k_{k1} \, \langle C_{K} \rangle  \, \delta C_{Ca} - k_{k1} \, \delta C_{Ca} \, \delta C_{K} + k_{d1} \delta C_{K-Ca} + k_{k1} \langle C_K \, C_{Ca} \rangle \\
&= \frac{d \langle C_{Ca} \rangle}{dt}  + I_{Ca} \, \delta p - k_l \, \delta C_{Ca} - k_{k1} \, \langle C_{K} \rangle  \,  \langle  C_{Ca} \rangle - k_{k1} \, \langle C_{Ca} \rangle  \, \delta C_{K} - k_{k1} \, \langle C_{K} \rangle  \, \delta C_{Ca} \\
& \quad - k_{k1} \, \delta C_{Ca} \, \delta C_{K}  + k_{d1} \delta C_{K-Ca} + k_{k1} \langle C_K \, C_{Ca} \rangle.
\end{aligned}
\label{rateCa}
\end{equation}

Now use the approximation to neglect the second order term involving fluctuations:
\begin{equation}
\begin{aligned}
\langle C_{Ca} \rangle \, \delta C_K + \langle C_{K} \rangle \, \delta C_{Ca} \gg \delta C_{Ca} \, \delta C_K.
\end{aligned}
\label{approx}
\end{equation}

Using Eq.~\ref{steady_state_values}, $\langle C_{K} \, C_{Ca}  \rangle$ can be written as:
\begin{equation}
\begin{aligned}
\langle C_{K} \, C_{Ca}  \rangle &= \langle C_{K} \rangle \, \langle C_{Ca} \rangle + \langle C_{K} \rangle \langle \delta C_{Ca} \rangle + \langle C_{Ca} \rangle \langle \delta C_{K} \rangle + \langle \delta C_{Ca} \, \delta C_{K} \rangle \approx \langle C_{K} \rangle \, \langle C_{Ca} \rangle,
\end{aligned}
\label{CkCca_expectation}
\end{equation}
where the second-order term involving fluctuations, \(\langle \delta C_{Ca} \, \delta C_{K} \rangle\), is neglected based on the linearization approximation (Eq.~\ref{approx}). Additionally, the conditions \( \langle \delta C_{Ca} \rangle = \langle \delta C_{K} \rangle = 0 \) follow from taking the expectation value on both sides of Eq.~\ref{steady_state_values}. Using Eqs. \ref{approx} and \ref{CkCca_expectation}, Eq. \ref{rateCa} can be written as:
\begin{equation}
\begin{aligned}
 \frac{d (\delta C_{Ca})}{dt} \approx I_{Ca} \, \delta p - k_l \, \delta C_{Ca} - k_{k1} \, \langle C_{Ca} \rangle  \, \delta C_{K} - k_{k1} \, \langle C_{K} \rangle  \, \delta C_{Ca} - k_{k1} \, \delta C_{Ca} \, \delta C_{K} + k_{d1} \delta C_{K-Ca}.
\end{aligned}
\end{equation}

In a similar manner, the time evolution of concentration fluctuations for \( C_{\text{K}} \), \( C_{\text{K-Ca}} \), and \( C_{\text{Pr}} \) can be derived. This results in the following linear rate equations:
\begin{align}
  \frac{d(\delta C_{Ca}(t))}{dt} &= I_{Ca} \, (p(t) - \bar{p}) - k_l \delta C_{Ca}(t) - k_{k1} \bar{C}_{\text{K}} \delta C_{Ca}(t) - k_{k1} \bar{C}_{\text{Ca}} \delta C_{K}(t) + k_{d1} \delta C_{K-Ca}(t)  \nonumber\\
  \frac{d(\delta C_{K}(t))}{dt} &= k_{d1} \delta C_{K-Ca}(t) - k_{k1} \bar{C}_{\text{K}} \delta C_{Ca}(t) \nonumber - k_{k1} \bar{C}_{\text{Ca}} \delta C_{K}(t) \nonumber \\
  \frac{d(\delta C_{K-Ca}(t))}{dt} &= k_{k1} \bar{C}_{\text{K}} \delta C_{Ca}(t) + k_{k1} \bar{C}_{\text{Ca}} \delta C_{K}(t) - k_{d1} \delta C_{K-Ca}(t)  \nonumber\\
  \frac{d(\delta C_{Pr}(t))}{dt} &= k_p \delta C_{K-Ca}(t) - k_{dp} \delta C_{Pr}(t).
\label{eq:linearized_eq}
\end{align}

\section{Solving Linearized Rate Equations}
\label{appendB}

To derive the expression for fluctuations in the phosphorylation readout, \( \delta C_{Pr}(t) \) by solving coupled linear ODEs (Eq.~\ref{eq:linearized_eq}), we apply the Laplace transform technique with the initial conditions \( \delta C_{\text{Ca}}(t=0) = \delta C_{\text{K}}(t=0) = \delta C_{\text{K-Ca}}(t=0) = \delta C_{\text{Pr}}(t=0) = 0 \). These initial conditions assume that at the measurement time \( t=0 \), the concentrations of the various cytoplasmic components are at their fixed initial values with no deviations. Some useful formulas for converting equations from the time domain (\( t \)) to the Laplace domain (\( s \)) are \cite{Arfken}:
\begin{align}
  F(s) &= \mathcal{L}[f(t)] = \int_{0}^{\infty} e^{-s \, t} \, f(t)  \, dt \nonumber \\
  \mathcal{L}[A] &= \frac{A}{s},  \quad \text{where $A$ is constant } \nonumber \\
  \mathcal{L}\left[\frac{df(t)}{dt} \right] &= s \, \mathcal{L}[f(t)] - f(t=0).
\end{align}
Using the above formulas and the notation \( C(t) \) to denote concentration in the time domain and \( C(s) \) for concentration in the Laplace domain, Eq.~\ref{eq:linearized_eq} transforms into the Laplace domain as follows:
\begin{align}
  s \, \delta C_{Ca}(s) - \delta C_{Ca}(t=0) &= I_{Ca} \left[p(s) - \frac{\bar{p}}{s} \right] - (k_l + k_{k1} \bar{C}_{\text{K}}) \delta C_{Ca}(s) - k_{k1} \bar{C}_{\text{Ca}} \delta C_{K}(s) + k_{d1} \delta C_{K-Ca}(s) \nonumber \\
  s \, \delta C_{K}(s) - \delta C_{K}(t=0) &= k_{d1} \delta C_{K-Ca}(s) - k_{k1} \bar{C}_{\text{K}} \delta C_{Ca}(s) - k_{k1} \bar{C}_{\text{Ca}} \delta C_{K}(s) \nonumber \\
  s \, \delta C_{K-Ca}(s) - \delta C_{K-Ca}(t=0) &= k_{k1} \bar{C}_{\text{K}} \delta C_{Ca}(s) + k_{k1} \bar{C}_{\text{Ca}} \delta C_{K}(s) - k_{d1} \delta C_{K-Ca}(s)  \nonumber\\
  s \, \delta C_{Pr}(s) - \delta C_{Pr}(t=0) &= k_p \delta C_{K-Ca}(s) - k_{dp} \delta C_{Pr}(s) \label{laplacerateeqn}.
\end{align}
Utilizing the above equations and the initial conditions  $\delta C_{K-Ca}(s)$, $\delta C_{K}(s)$, and $\delta C_{Ca}(s)$ can be expressed in terms of $\delta C_{Pr}(s)$ as follows:
\begin{align}
\delta C_{K-Ca}(s) &= \frac{s + k_{dp}}{k_p} \, \delta C_{Pr}(s) \label{laplaceCkca} \\
\delta C_{K}(s) &= - \delta C_{K-Ca}(s) = - \bigg( \frac{s + k_{dp}}{k_p} \bigg) \, \delta C_{Pr}(s)
 \label{laplaceCk} \\
\delta C_{Ca}(s) &= \frac{1}{k_{k1} \bar{C}_{\text{K}}} \left[(s + k_{d1}) \delta C_{K-Ca}(s) - k_{k1} \bar{C}_{\text{Ca}} \delta C_{K}(s) \right] \nonumber \\
&= \frac{1}{k_{k1} \bar{C}_{\text{K}}} \bigg[(s + k_{d1}) \bigg(\frac{s + k_{dp}}{k_p} \bigg) + k_{k1} \bar{C}_{\text{Ca}} \bigg(\frac{s + k_{dp}}{k_p} \bigg) \bigg] \, \delta C_{Pr}(s). 
\label{laplaceCa}
\end{align} 
Using Eqs. \ref{laplaceCkca}, \ref{laplaceCk}, and \ref {laplaceCa}, Eq. \ref{laplacerateeqn} gives:
\begin{equation}
\begin{aligned}
I_{Ca} \, \bigg[ p(s) - \frac{\Bar{p}}{s} \bigg]  &= (s + k_l + k_{k1} \bar{C}_{\text{K}}) \delta C_{Ca}(s) + k_{k1} \bar{C}_{\text{Ca}} \delta C_{K}(s) - k_{d1} \delta C_{K-Ca}(s) \\
&= \frac{(s + k_l + k_{k1} \bar{C}_{\text{K}})}{k_{k1} \bar{C}_{\text{K}}} \bigg[(s + k_{d1}) \bigg(\frac{s + k_{dp}}{k_p} \bigg) + k_{k1} \bar{C}_{\text{Ca}} \bigg(\frac{s + k_{dp}}{k_p} \bigg) \bigg] \, \delta C_{Pr}(s) \\
& \quad - k_{k1} \bar{C}_{\text{Ca}} \bigg( \frac{s + k_{dp}}{k_p} \bigg) \, \delta C_{Pr}(s) - k_{d1} \, \bigg( \frac{s + k_{dp}}{k_p} \bigg) \, \delta C_{Pr}(s) \\
&= \bigg[ \frac{(s + k_l + k_{k1} \bar{C}_{\text{K}})}{k_{k1} \bar{C}_{\text{K}}} \bigg((s + k_{d1}) \bigg(\frac{s + k_{dp}}{k_p} \bigg) + k_{k1} \bar{C}_{\text{Ca}} \bigg(\frac{s + k_{dp}}{k_p} \bigg) \bigg) \\
& \quad - k_{k1} \bar{C}_{\text{Ca}} \bigg( \frac{s + k_{dp}}{k_p} \bigg)  - k_{d1} \, \bigg( \frac{s + k_{dp}}{k_p} \bigg) \bigg] \, \delta C_{Pr}(s).
\end{aligned} \label{laplacecpr1}
\end{equation}
Above equation can be expressed more compactly as:
\begin{equation}
\begin{aligned}
\delta C_{Pr}(s) &= \frac{I_{Ca}}{A(s)} \, \bigg[ p(s) - \frac{\Bar{p}}{s} \bigg],
\end{aligned} \label{laplacecpr}
\end{equation}
where the function \( A(s) \) is given by
\begin{equation}
\begin{aligned}
A(s) &=  \frac{(s + k_l + k_{k1} \bar{C}_{\text{K}})}{k_{k1} \bar{C}_{\text{K}}} \bigg[(s + k_{d1}) \bigg(\frac{s + k_{dp}}{k_p} \bigg) + k_{k1} \bar{C}_{\text{Ca}} \bigg(\frac{s + k_{dp}}{k_p} \bigg) \bigg] \\
& \quad - k_{k1} \bar{C}_{\text{Ca}} \bigg( \frac{s + k_{dp}}{k_p} \bigg) - k_{d1} \, \bigg( \frac{s + k_{dp}}{k_p} \bigg) \\
&= \frac{1}{\bar{C}_{\text{K}} k_{k1} k_{p}} \, s^3 + \bigg( \frac{{1}}{{k_{p}}} + \frac{{\bar{C}_{\text{Ca}}}}{{\bar{C}_{\text{K}} k_{p}}} + \frac{{k_{d1}}}{{\bar{C}_{\text{K}} k_{k1} k_{p}}} + \frac{{k_{dp}}}{{\bar{C}_{\text{K}} k_{k1} k_{p}}} + \frac{{k_{l}}}{{\bar{C}_{\text{K}} k_{k1} k_{p}}} \bigg) \, s^2 \\
& + \bigg(\frac{{k_{dp}}}{{k_{p}}} + \frac{{\bar{C}_{\text{Ca}} k_{dp}}}{{\bar{C}_{\text{K}} k_{p}}} + \frac{{k_{d1} k_{dp}}}{{\bar{C}_{\text{K}} k_{k1} k_{p}}} + \frac{{\bar{C}_{\text{Ca}} k_{l}}}{{\bar{C}_{\text{K}} k_{p}}} + \frac{{k_{d1} k_{l}}}{{C_{k} k_{k1} k_{p}}} + \frac{{k_{dp} k_{l}}}{{\bar{C}_{\text{K}} k_{k1} k_{p}}} \bigg) \, s^1 + \bigg( \frac{{\bar{C}_{\text{Ca}} k_{dp} k_{l}}}{{\bar{C}_{\text{K}} k_{p}}} + \frac{{k_{d1} k_{dp} k_{l}}}{{\bar{C}_{\text{K}} k_{k1} k_{p}}}\bigg) \, s^0 \\
&= A_3 \, s^3 + A_2 \, s^2 + A_1 \, s^1 + A_0 \, s^0 \\
&= \sum_{n=0}^{3}A_n \, s^{n}.
\end{aligned}
\label{laplacecoeff}
\end{equation}
The variables $A_0$, $A_1$, $A_2$, and $A_3$ introduced in above equation are given by
\begin{equation}
\begin{aligned}
A_0 &= \frac{{\bar{C}_{\text{Ca}} k_{dp} k_{l}}}{{\bar{C}_{\text{K}} k_{p}}} + \frac{{k_{d1} k_{dp} k_{l}}}{{\bar{C}_{\text{K}} k_{k1} k_{p}}} \\
A_1 &= \frac{{k_{dp}}}{{k_{p}}} + \frac{{\bar{C}_{\text{Ca}} k_{dp}}}{{\bar{C}_{\text{K}} k_{p}}} + \frac{{k_{d1} k_{dp}}}{{\bar{C}_{\text{K}} k_{k1} k_{p}}} + \frac{{\bar{C}_{\text{Ca}} k_{l}}}{{\bar{C}_{\text{K}} k_{p}}} + \frac{{k_{d1} k_{l}}}{{C_{k} k_{k1} k_{p}}} + \frac{{k_{dp} k_{l}}}{{\bar{C}_{\text{K}} k_{k1} k_{p}}} \\
A_2 &= \frac{{1}}{{k_{p}}} + \frac{{\bar{C}_{\text{Ca}}}}{{\bar{C}_{\text{K}} k_{p}}} + \frac{{k_{d1}}}{{\bar{C}_{\text{K}} k_{k1} k_{p}}} + \frac{{k_{dp}}}{{\bar{C}_{\text{K}} k_{k1} k_{p}}} + \frac{{k_{l}}}{{\bar{C}_{\text{K}} k_{k1} k_{p}}} \\
A_3 &= \frac{1}{\bar{C}_{\text{K}} k_{k1} k_{p}}.
\end{aligned}
\label{laplacecoeffvalue}
\end{equation}

\( \delta C_{Pr}(t) \) in the time domain is obtained by applying the inverse Laplace transform to both sides of Eq. \ref{laplacecpr}, as:
\begin{equation}
\begin{aligned}
\delta C_{Pr}(t) = \mathcal{L}^{-1}\{\delta C_{Pr}(s)\} = I_{\text{Ca}} \, \mathcal{L}^{-1} \left\{ \frac{1}{A(s)} \, p(s) \right\} - I_{\text{Ca}} \, \Bar{p} \, \mathcal{L}^{-1} \left\{ \frac{1}{A(s)} \, \frac{1}{s} \right\}.
\end{aligned}
\end{equation}
Utilizing the convolution theorem, which states that $\mathcal{L}^{-1} \left\{ f_1(s) \, f_2(s) \right\} = \mathcal{L}^{-1} \left\{ f_1(s) \right\} \ast \mathcal{L}^{-1} \left\{ f_2(s) \right\} = f_1(t) \ast f_2(t) = \int_{0}^{t} f_1(t - t') f_2(t') \, dt'$ \cite{Arfken}, along with the fact that \(\mathcal{L}^{-1} \left\{ \frac{1}{s} \right\} = 1\), and using the notations
\(I_{\text{Ca}} \mathcal{L}^{-1} \left\{ \frac{1}{A(s)} \right\} = f(t)\) and  \( \mathcal{L}^{-1} \left\{ p(s) \right\} = p(t)\), above equation at the measurement time \( t = T \) results in:
\begin{equation}
\begin{aligned}
\delta C_{Pr}(T) &= \int_{0}^{T} f(T-t') \, p(t') \, dt' - \, \Bar{p} \, \int_{0}^{T} f(T-t') \, dt'.
\end{aligned}
\label{convolutioncprT}
\end{equation}
which can be written as:
\begin{equation}
\begin{aligned}
\delta C_{Pr}(T) &= \int_{0}^{T} f(T - t') \, \delta p(t') \, dt'.
\end{aligned}
\end{equation}

\section{RMS Error in Inferring the Mean Ligand Concentration}
\label{appendC}

The RMS error in estimating the mean ligand concentration once the cell has reached steady state, as given by Eq. 6 in the main text, is derived from the phosphorylation readout. To obtain this, we first compute the variance in the phosphorylation readout, denoted by \( \sigma_{C_{\text{Pr}}}^2(T) \), at the measurement time \( T \), using the fluctuations in the phosphorylation readout as expressed by Eq. \ref{convolutioncprT}. Consider an ensemble of a large number of identical copies of the system. The expectation value of the fluctuation in the phosphorylation readout is given by:
\begin{equation}
\begin{aligned}
\langle \delta C_{Pr}(T) \rangle &=  \int_{0}^{T} f(T-t') \, \langle p(t') \rangle \, dt' - \Bar{p} \, \int_{0}^{T} f(T-t') \, dt' \\
&= \Bar{p} \, \int_{0}^{T} f(T-t') \, dt' -  \Bar{p} \, \int_{0}^{T} f(T-t') \, dt' \\
&= 0.
\end{aligned}
\label{delcpravg}
\end{equation}
Note that the expectation applies only to the occupation probability of the receptor \( p(t) \), as it is stochastic in nature. In the first term of the above equation, \( \langle p(t) \rangle = \bar{p} \) is used. The square fluctuation, denoted by $\delta C_{Pr}(T)^2$, is calculated using Eq. \ref{convolutioncprT} as:
\begin{equation}
\begin{aligned}
& \delta C_{Pr}(T)^2 \\
&= \bigg[\int_{t'=0}^{T} f(T-t') \, p(t') \, dt' - \Bar{p} \, \int_{t'=0}^{T} f(T-t') \, dt' \bigg] \cdot \bigg[ \int_{t''=0}^{T} f(T-t'') \, p(t'') \, dt'' - \Bar{p} \, \int_{t''=0}^{T} f(T-t'') \, dt'' \bigg] \\
&= \bigg[\int_{t'=0}^{T} \int_{t''=0}^{T} f(T-t') \, f(T-t'') \,  p(t') \, p(t'') \, dt' \, dt'' + \Bar{p}^2 \, \int_{t'=0}^{T} \int_{t''=0}^{T} f(T-t') \, f(T-t'') \, dt' \, dt'' \\
& \quad - \Bar{p} \, \int_{t'=0}^{T} \int_{t''=0}^{T} f(T-t') \, f(T-t'') \, p(t') \, dt' \, dt'' - \Bar{p} \, \int_{t'=0}^{T} \int_{t''=0}^{T} f(T-t') \, f(T-t'') \, p(t'') \, dt' \, dt''  \bigg].
\end{aligned} \label{delcprsqr}
\end{equation}
The Mean square fluctuation, denoted by $\langle \delta C_{Pr}(T)^2 \rangle$, is calculated as:
\begin{equation}
\begin{aligned}
& \langle \delta C_{Pr}(T)^2 \rangle \\
&= \bigg[\int_{t'=0}^{T} \int_{t''=0}^{T} f(T-t') \, f(T-t'') \, \langle  p(t') \, p(t'') \rangle \, dt' \, dt'' + \Bar{p}^2 \, \int_{t'=0}^{T} \int_{t''=0}^{T} f(T-t') \, f(T-t'') \, dt' \, dt'' \\
& \quad - \Bar{p} \, \int_{t'=0}^{T} \int_{t''=0}^{T} f(T-t') \, f(T-t'') \, \langle  p(t') \rangle  \, dt' \, dt'' - \Bar{p} \, \int_{t'=0}^{T} \int_{t''=0}^{T} f(T-t') \, f(T-t'') \, \langle p(t'') \rangle \, dt' \, dt''  \bigg] \\
&=  \bigg[\int_{t'=0}^{T} \int_{t''=0}^{T} f(T-t') \, f(T-t'') \, \langle  p(t') \, p(t'') \rangle \, dt' \, dt'' + \bigg( \Bar{p} \, \int_{t=0}^{T} f(T-t) \, dt \bigg)^2 - 2 \, \bigg( \Bar{p} \, \int_{t=0}^{T} f(T-t) \, dt \bigg)^2 \bigg] \\
&= \bigg[\int_{t'=0}^{T} \int_{t''=0}^{T} f(T-t') \, f(T-t'') \, \langle  p(t') \, p(t'') \rangle \, dt' \, dt'' - \bigg( \Bar{p} \, \int_{t=0}^{T} f(T-t) \, dt \bigg)^2 \bigg],
\end{aligned} \label{delcprsqravg}
\end{equation}
where \( \langle p(t') \, p(t'') \rangle \) represents the autocorrelation function of \( p(t) \), which can be taken as an exponentially decaying function, as described in Ref. \cite{Berg1977}:
\begin{equation}
G(t', t'') = \langle p(t') \, p(t'') \rangle = \bar{p}^2 + \bar{p} \, (1-\bar{p}) \, e^{-\frac{|t''-t'|}{(1-\bar{p}) \tau_b}},
\label{autocorrelationberg}
\end{equation}
where \( \bar{p} \) is the average occupation probability of a receptor, and \( (1 - \bar{p}) \tau_b = \tau_c \) is the correlation time, representing how long the receptor’s state (occupied or unoccupied) remains correlated with its past state. Here, \( \tau_b = k_{-}^{-1} \) goes as the average time a ligand remains bound to a receptor. Hence Eq. \ref{delcprsqravg} becomes:
\begin{equation}
\begin{aligned}
& \langle \delta C_{Pr}(T)^2 \rangle \\
&=  \Bar{p}^2 \int_{t'=0}^{T} \int_{t''=0}^{T} f(T-t') \, f(T-t'') \, dt' \, dt''
+ \bar{p} \, (1-\bar{p}) \int_{t'=0}^{T} \int_{t''=0}^{T} f(T-t') \, f(T-t'') \,  e^{-\frac{|t''-t'|}{(1-\bar{p}) \tau_b}} \, dt' \, dt'' \\
& \quad - \bigg( \Bar{p} \, \int_{t=0}^{T} f(T-t) \, dt \bigg)^2  \\
&=  \bigg( \Bar{p} \, \int_{t=0}^{T} f(T-t) \, dt \bigg)^2 + \bar{p} \, (1-\bar{p}) \int_{t'=0}^{T} \int_{t''=0}^{T} f(T-t') \, f(T-t'') \,  e^{-\frac{|t''-t'|}{(1-\bar{p}) \tau_b}} \, dt' \, dt'' - \bigg( \Bar{p} \, \int_{t=0}^{T} f(T-t) \, dt \bigg)^2  \\
&= \bar{p} \, (1-\bar{p}) \int_{t'=0}^{T} \int_{t''=0}^{T} f(T-t') \, f(T-t'') \,  e^{-\frac{|t''-t'|}{(1-\bar{p}) \tau_b}} \, dt' \, dt'' .
\end{aligned} 
\label{delcprsqravg1}
\end{equation}
The variance in the phosphorylation readout, using Eqs. \ref{delcpravg} and \ref{delcprsqravg1}, is expressed as:
\begin{equation}
\begin{aligned}
\sigma_{{C}_{\text{Pr}}}^2(T) = \langle \delta C_{Pr}(T)^2 \rangle -  \langle \delta C_{Pr}(T) \rangle^2 = \bar{p} \, (1-\bar{p}) \int_{t'=0}^{T} \int_{t''=0}^{T} f(T-t') \, f(T-t'') \,  e^{-\frac{|t''-t'|}{(1-\bar{p}) \tau_b}} \, dt' \, dt'' .
\end{aligned} 
\end{equation}
In the limit of large measurement times, \( \lim_{T \to \infty} \), the above equation can be rewritten as:
\begin{equation}
\begin{aligned}
\sigma_{{C}_{\text{Pr}}}^2 = \lim_{T \to \infty} \langle \delta C_{Pr}(T)^2 \rangle &= \lim_{T \to \infty}  \, \bigg[\bar{p} \, (1-\bar{p}) \int_{t'=0}^{T} \int_{t''=0}^{T} f(T-t') \, f(T-t'') \,  e^{-\frac{|t''-t'|}{(1-\bar{p}) \tau_b}} \, dt' \, dt'' \bigg].
\end{aligned}
\label{varincpr}
\end{equation}
The function \( f(t) \) used in above equation is the inverse Laplace transform of \( I_{\text{Ca}}/A(s) \). From Eq. \ref{laplacecoeff}, $A(s)$ is cubic polynomial in $s$. Thus, \( A(s) = \sum_{n=0}^{3} A_n \, s^{n} = A_3 \, (s - s_1) \, (s - s_2) \, (s - s_3) \), where \( A_n = \{A_0, A_1, A_2, A_3\} \) are defined in Eq. \ref{laplacecoeffvalue} and \( s_k \) represent the roots of the cubic equation \( A(s) = \sum_{n=0}^{3} A_n \, s^{n} = 0 \), solved for \( s \) and given by:
\begin{equation}
\begin{aligned}
s_1 &= \frac{1}{2} \left( -k_{d1} - \Bar{C}_{Ca} k_{k1} - \Bar{C}_{K} k_{k1} - k_{l} - \sqrt{(k_{d1} + \Bar{C}_{Ca} k_{k1} + \Bar{C}_{K} k_{k1} + k_{l})^2 - 4 (k_{d1} k_{l} + \Bar{C}_{Ca} k_{k1} k_{l})} \right) \\
s_2 &= \frac{1}{2} \left( -k_{d1} - \Bar{C}_{Ca} k_{k1} - \Bar{C}_{K} k_{k1} - k_{l} + \sqrt{(k_{d1} + \Bar{C}_{Ca} k_{k1} + \Bar{C}_{K} k_{k1} + k_{l})^2 - 4 (k_{d1} k_{l} + \Bar{C}_{Ca} k_{k1} k_{l})} \right)  \\
s_3 &= -k_{dp}.
\end{aligned} 
\label{svalues}
\end{equation}

\( 1/A(s) \) can be written as a sum of partial fractions, as follows:
\begin{equation}
\begin{aligned}
\frac{1}{A(s)} &= \frac{1}{A_3 (s - s_1)(s - s_2)(s - s_3)}\nonumber\\
&= \frac{1}{A_3 (s_1 - s_2)(s_1 - s_3)} \cdot \frac{1}{s - s_1} + \frac{1}{A_3 (s_2 - s_1)(s_2 - s_3)} \cdot \frac{1}{s - s_2} + \frac{1}{A_3 (s_3 - s_1)(s_3 - s_2)} \cdot \frac{1}{s - s_3}.
\end{aligned} 
\end{equation}
Taking the inverse Laplace transform of each term and applying the property \(\mathcal{L}^{-1}\left( \frac{1}{s - a} \right) = e^{at}\), yields:
\begin{equation}
\begin{aligned}
f(t) = \frac{I_{\text{Ca}}}{A_3 (s_1 - s_2)(s_1 - s_3)} e^{s_1 t} + \frac{I_{\text{Ca}}}{A_3 (s_2 - s_1)(s_2 - s_3)} e^{s_2 t} + \frac{I_{\text{Ca}}}{A_3 (s_3 - s_1)(s_3 - s_2)} e^{s_3 t},
\end{aligned} 
\end{equation}
which can be written as:
\begin{equation}
\begin{aligned}
f(t) = I_{\text{Ca}}\sum_{k=1}^{3}  \bigg[ \frac{e^{s_k \, t}}{A_3\prod_{\substack{l=1 \\ l \neq k}}^{3} (s_k - s_l)} \bigg] = I_{\text{Ca}}\bar{C}_{K} k_{k1} k_{p} \, \sum_{k=1}^{3}  \bigg[ \frac{e^{s_k \, t}}{\prod_{\substack{l=1 \\ l \neq k}}^{3} (s_k - s_l)} \bigg].
\end{aligned}
\label{weigting_function}
\end{equation}
Substituting above expression for $f(t)$ back into Eq. \ref{varincpr} gives
\begin{equation}
\begin{aligned}
& \sigma_{{C}_{\text{Pr}}}^2 \\
&= \lim_{T \to \infty} I_{\text{Ca}}^2 \, \bigg[\bar{p} \, (1-\bar{p}) \int_{t'=0}^{T} \int_{t''=0}^{T} \bigg(\sum_{k=1}^{3}  \frac{e^{s_k \, (T-t')}}{A_3\prod_{\substack{l=1 \\ l \neq k}}^{3} (s_k - s_l)} \bigg) \, \bigg(\sum_{m=1}^{3}  \frac{e^{s_m \, (T-t'')}}{A_3\prod_{\substack{n=1 \\ n \neq m}}^{3} (s_m - s_n)} \bigg) \,  e^{-\frac{|t''-t'|}{(1-\bar{p}) \tau_b}} \, dt' \, dt'' \bigg] \\
&=\frac{ I_{\text{Ca}}^2 \, \bar{p} \, (1-\bar{p}) }{A_3^2}\, \sum_{k=1}^{3} \, \sum_{m=1}^{3} \, \frac{1}{ \prod_{\substack{l=1 \\ l \neq k}}^{3} (s_k - s_l) \cdot \prod_{\substack{n=1 \\ n \neq m}}^{3} (s_m - s_n) } \,  \lim_{T \to \infty} \, \bigg(e^{(s_{k}+s_{m}) \, T} \, \int_{t'=0}^{T} \, dt' \,  e^{-s_{k} \, t'} \, \int_{t''=0}^{T} \, e^{-s_{m} \, t''} \, e^{-\frac{|t'-t''|}{(1-\bar{p}) \tau_b}} \,  dt'' \bigg) \\
&= \frac{I_{\text{Ca}}^2 }{A_3^2}\, \sum_{k=1}^{3} \, \sum_{m=1}^{3} \, \left[  \frac{1}{ \prod_{\substack{l=1 \\ l \neq k}}^{3} (s_k - s_l) \cdot \prod_{\substack{n=1 \\ n \neq m}}^{3} (s_m - s_n) } \, \left( \frac{\bar{p} \, (1 - \bar{p})^2 \, \tau_b \, \left[ (1 - \bar{p}) \, \tau_b - \frac{2}{s_{k} + s_{m}}\right] }{\left[1 - (1 - \bar{p}) \, \tau_b\, s_{k}\right] \, \left[1 - (1 - \bar{p}) \, \tau_b \, s_{m} \right]} \right) \right].
\end{aligned} 
\label{cprvarcontour}
\end{equation}
Substituting \( s_1 \), \( s_2 \), and \( s_3 \) from Eq. \ref{svalues} into the above equation, and simplifying, yields:
\begin{equation}
\begin{aligned}
\sigma_{{C}_{\text{Pr}}}^2 &= \frac{16 \, I_{\text{Ca}}^2 \, \bar{p} \, (1-\bar{p}) \, \tau_c}{A_3^2}\, \bigg[ \frac{4 (a + d) + 4 \, (a + d)^2 \, \tau_c + a \, [(a + 2d)^2-b] \, \tau_c^2}{a \, (a^2 - b) \, d \, [(a + 2d)^2-b] \, (1 + d \tau_c) \, [4 + 4a \tau_c + (a^2 - b) \tau_c^2]} \bigg],
\end{aligned} 
\label{finalcprvariance}
\end{equation}
with various parameters given by:
\begin{equation}
\begin{aligned}
\tau_c &= (1 - \bar{p}) \, \tau_b \\
\tau_b &= \frac{1}{k_-} \\
\bar{p} &= \frac{k_+ \, \bar{c}}{k_+ \, \bar{c} + k_-} \\
a &= k_{d1} + \Bar{C}_{Ca} k_{k1} + \Bar{C}_{K} k_{k1} + k_{l} \\
b &= a^2 - 4 (k_{d1} k_{l} + \Bar{C}_{Ca} k_{k1} k_{l}) \\
d &= k_{dp} \\
\bar{C}_{Ca} &= \frac{I_{Ca} \, \bar{p} }{k_l} \\
\bar{C}_{K} &= \frac{C_{K0} \, k_{d1} \, k_l}{k_{d1} \, k_l + I_{Ca} \, k_{k1} \, \bar{p}}.
\end{aligned}
\label{parameters_rms}
\end{equation}
The expression for $A_3$ is provided in Eq.~\ref{laplacecoeffvalue}, while the mean concentrations and the average occupation probability are derived in Appendix~\ref{appendF}.

RMS error in sensing the ligand concentration is calculated using the relation $\frac{\delta c}{c} = \frac{\delta C_{\text{Pr}}}{c \left( \frac{\partial C_{\text{Pr}}}{\partial c} \right)}$, which translates to (see Appendix~\ref{appendG} for detailed derivation):
\begin{equation}
\begin{aligned}
\frac{\Delta c_{rms}}{\bar{c}} = \frac{{k_{dp} \, (k_{d1} \, k_l + I_{\text{ca}} \, k_{k1} \, \bar{p})^2}}{{C_{k0} \, I_{\text{ca}} k_{d1} \, k_{k1} \, k_l \, k_p \, \bar{p} \, (1 - \bar{p})}} \, \sigma_{C_{\text{Pr}}}.
\end{aligned}
\label{errorinconc2}
\end{equation}
Substituting Eq. \ref{finalcprvariance} into above equation gives:
\begin{equation}
\begin{aligned}
&\frac{\Delta c_{rms}}{\bar{c}} \\
&= \bigg[ \frac{{k_{dp} \, (k_{d1} \, k_l + I_{\text{ca}} \, k_{k1} \, \bar{p})^2}}{{C_{k0} \, I_{\text{ca}} k_{d1} \, k_{k1} \, k_l \, k_p \, \bar{p} \, (1 - \bar{p})}} \frac{1}{A_3}\bigg] \, \bigg[16 \, I_{\text{Ca}}^2 \, \bar{p} \, (1-\bar{p}) \, \tau_c \, \frac{4 (a + d) + 4 \, (a + d)^2 \, \tau_c + a \, [(a + 2d)^2-b] \, \tau_c^2}{a \, (a^2 - b) \, d \, [(a + 2d)^2-b] \, (1 + d \tau_c) \, [4 + 4a \tau_c + (a^2 - b) \tau_c^2]} \bigg]^{\frac{1}{2}} \\
&= \bigg[\frac{{4 \, k_{dp} \, (k_{d1} \, k_l + I_{\text{ca}} \, k_{k1} \, \bar{p})^2}}{{C_{k0} \, k_{d1} \, k_{k1} \, k_l \, k_p}}\frac{1}{A_3} \bigg] \, \bigg[ \frac{ \tau_c}{\bar{p} \, (1 - \bar{p})} \, \frac{4 (a + d) + 4 \, (a + d)^2 \, \tau_c + a \, [(a + 2d)^2-b] \, \tau_c^2}{a \, (a^2 - b) \, d \, [(a + 2d)^2-b] \, (1 + d \tau_c) \, [4 + 4a \tau_c + (a^2 - b) \tau_c^2]} \bigg]^{\frac{1}{2}} \\
&=\bigg[\frac{{4 \, k_{dp} \, (k_{d1} \, k_l + I_{\text{ca}} \, k_{k1} \, \bar{p})}}{{\bar{C_{k}} \, k_{k1}\, k_p}}\frac{1}{A_3} \bigg] \, \bigg[ \frac{ \tau_c}{\bar{p} \, (1 - \bar{p})} \, \frac{4 (a + d) + 4 \, (a + d)^2 \, \tau_c + a \, [(a + 2d)^2-b] \, \tau_c^2}{a \, (a^2 - b) \, d \, [(a + 2d)^2-b] \, (1 + d \tau_c) \, [4 + 4a \tau_c + (a^2 - b) \tau_c^2]} \bigg]^{\frac{1}{2}}\\
&= \bigg[{4 \, k_{dp} \, (k_{d1} \, k_l + I_{\text{ca}} \, k_{k1} \, \bar{p})A_3}\cdot\frac{1}{A_3} \bigg] \, \bigg[ \frac{ \tau_c}{\bar{p} \, (1 - \bar{p})} \, \frac{4 (a + d) + 4 \, (a + d)^2 \, \tau_c + a \, [(a + 2d)^2-b] \, \tau_c^2}{a \, (a^2 - b) \, d \, [(a + 2d)^2-b] \, (1 + d \tau_c) \, [4 + 4a \tau_c + (a^2 - b) \tau_c^2]} \bigg]^{\frac{1}{2}}\\
&= \bigg[4 \, k_{dp} \, (k_{d1} \, k_l +\, k_{k1}\Bar{C}_{Ca}\, k_l) \bigg] \, \bigg[ \frac{ \tau_c}{\bar{p} \, (1 - \bar{p})} \, \frac{4 (a + d) + 4 \, (a + d)^2 \, \tau_c + a \, [(a + 2d)^2-b] \, \tau_c^2}{a \, (a^2 - b) \, d \, [(a + 2d)^2-b] \, (1 + d \tau_c) \, [4 + 4a \tau_c + (a^2 - b) \tau_c^2]} \bigg]^{\frac{1}{2}}\\
&= \bigg[d(a^2-b) \bigg] \, \bigg[ \frac{ \tau_c}{\bar{p} \, (1 - \bar{p})} \, \frac{4 (a + d) + 4 \, (a + d)^2 \, \tau_c + a \, [(a + 2d)^2-b] \, \tau_c^2}{a \, (a^2 - b) \, d \, [(a + 2d)^2-b] \, (1 + d \tau_c) \, [4 + 4a \tau_c + (a^2 - b) \tau_c^2]} \bigg]^{\frac{1}{2}},
\end{aligned}
\end{equation}
which can be simplified as:
\begin{equation}
\begin{aligned}
\frac{\Delta c_{rms}}{\bar{c}} =  \, \bigg[ \frac{ d(a^2-b) \tau_c}{\bar{p} \, (1 - \bar{p})} \, \frac{4 (a + d) (1+ (a + d) \, \tau_c) + a \, [(a + 2d)^2-b] \, \tau_c^2}{a  \, [(a + 2d)^2-b] \, (1 + d \tau_c) \, [4 + 4a \tau_c + (a^2 - b) \tau_c^2]} \bigg]^{\frac{1}{2}},
\end{aligned}
\label{errorinconc3}
\end{equation}
where the various parameters are defined in Eq.~\ref{parameters_rms}.

\section{Lower Bound for RMS Error in Sensing Ligand Concentration}
\label{appendD}

The lower bound of the RMS error in sensing ligand concentration (Eq. 11) can be derived by minimizing the numerator and maximizing the denominator of Eq.~\ref{errorinconc3} through term reduction. The equal sign (\(=\)) is used to indicate exact simplifications, while the greater than sign (\(>\)) is applied when terms are neglected to obtain the lower bound. This process yields:

\begin{equation}
\begin{aligned}
\frac{\Delta c_{rms}}{\bar{c}} &>  \, \bigg[ \frac{ d(a^2-b) \tau_c}{\bar{p} \, (1 - \bar{p})} \, \frac{ a \, [(a + 2d)^2-b] \, \tau_c^2}{a  \, [(a + 2d)^2-b] \, (1 + d \tau_c) \, [4 + 4a \tau_c + (a^2 - b) \tau_c^2]} \bigg]^{\frac{1}{2}}\\
&= \, \bigg[ \frac{ d(a^2-b) \tau_c}{\bar{p} \, (1 - \bar{p})} \, \frac{  \tau_c^2}{ \, (1 + d \tau_c) \, [4 + 4a \tau_c + (a^2 - b) \tau_c^2]} \bigg]^{\frac{1}{2}}\\
&> \, \bigg[ \frac{ d(a^2-b) \tau_c}{\bar{p} \, (1 - \bar{p})} \, \frac{  \tau_c^2}{ \, (1 + d \tau_c) \, [4 + 4a \tau_c + a^2  \tau_c^2]} \bigg]^{\frac{1}{2}} \\
&= \, \bigg[ \frac{ d(a^2-b) \tau_c}{\bar{p} \, (1 - \bar{p})} \, \frac{  \tau_c^2}{ \, (1 + d \tau_c) \, ( a\tau_c+2)^2} \bigg]^{\frac{1}{2}}\\
&= \, \bigg[ \frac{ d \tau_c}{ (1 + d \tau_c)\,} \frac{(a^2-b)}{ \bar{p} (1 - \bar{p})}\,\bigg] ^{\frac{1}{2}}\frac{  \tau_c}{ \, \, ( a\tau_c+2)}.
\end{aligned}
\end{equation}
Substituting the parameters defined in Eq.~\ref{parameters_rms} into the above inequality yields:
\begin{equation}
\begin{aligned}
\frac{\Delta c_{rms}}{\bar{c}} 
&> \, \left[ \frac{  \frac{k_{dp}}{k_+ \, \bar{c} + k_-}}{ 1 +\frac{k_{dp}}{k_+ \, \bar{c} + k_-}} \frac{4 (k_{d1} k_{l} + \Bar{C}_{Ca} k_{k1} k_{l})}{ \frac{k_+ \, \bar{c}k_-}{(k_+ \, \bar{c} + k_-)^2} }\right] ^{\frac{1}{2}}\frac{ \frac{1}{k_+ \, \bar{c} + k_-}}{ (k_{d1} + \Bar{C}_{Ca} k_{k1} + \Bar{C}_{K} k_{k1} + k_{l}) \frac{1}{k_+ \, \bar{c} + k_-}+ 2}\\
&= \, \left[ \frac{ k_{dp}}{ (k_+ \, \bar{c} + k_- )+k_{dp}} \frac{4 (k_{d1} k_{l} + I_{Ca}\Bar{p}\, k_{k1})}{ \frac{k_+ \, \bar{c}k_-}{(k_+ \, \bar{c} + k_-)^2} }\right] ^{\frac{1}{2}}\frac{ 1}{ (k_{d1} + \Bar{C}_{Ca} k_{k1} + \Bar{C}_{K} k_{k1} + k_{l}) + 2(k_+ \, \bar{c} + k_-)}\\
&>\, \left[ \frac{ k_{dp}}{ (k_+ \, \bar{c} + k_- )+k_{dp}} \frac{4  I_{Ca}\Bar{p}\, k_{k1} }{ \frac{k_+ \, \bar{c}k_-}{(k_+ \, \bar{c} + k_-)^2} }\right] ^{\frac{1}{2}}\frac{ 1}{ (k_{d1} + \Bar{C}_{Ca} k_{k1} + \Bar{C}_{K} k_{k1} + k_{l}) + 2(k_+ \, \bar{c} + k_-)}\\
&=\, \left[ \frac{ k_{dp}}{ (k_+ \, \bar{c} + k_- )+k_{dp}} \frac{4  I_{Ca} k_{k1} }{ \frac{k_-}{(k_+ \, \bar{c} + k_-)} }\right] ^{\frac{1}{2}}\frac{ 1}{ k_{d1} + \frac{I_{\text{Ca}} \, \bar{p} \, k_{k1}}{k_l} + \frac{C_{k0} \, k_{d1} \, k_{k1} \, k_l}{k_{d1} \, k_l + I_{\text{Ca}} \, k_{k1} \, \bar{p}} + k_{l} + 2(k_+ \, \bar{c} + k_-)}\\
&>\, \left[ \frac{ k_{dp}}{ (k_+ \, \bar{c} + k_- )+k_{dp}} \frac{4  I_{Ca} k_{k1} }{ \frac{k_-}{(k_+ \, \bar{c} + k_-)} }\right] ^{\frac{1}{2}}\frac{ 1}{ k_{d1} + \frac{I_{\text{Ca}} \,  \, k_{k1}}{k_l} + C_{k0} \, k_{k1} + k_{l} + 2(k_+ \, \bar{c} + k_-)}\\
&=\, \left[ \frac{ k_{dp}}{ (k_+ \, \bar{c} + k_- )+k_{dp}} \frac{(k_+c+k_-)4  I_{Ca} k_{k1} }{ k_- }\right] ^{\frac{1}{2}}\frac{ 1}{ k_{d1} + \frac{I_{\text{Ca}} \,  \, k_{k1}}{k_l} + C_{k0} \, k_{k1} + k_{l} + 2(k_+ \, \bar{c} + k_-)}\\
&>\, \left[ \frac{4  I_{Ca} k_{k1} k_{dp}}{ (k_+ \, \bar{c} + k_- )+k_{dp}}  \right] ^{\frac{1}{2}}\frac{ 1}{ k_{d1} + \frac{I_{\text{Ca}} \,  \, k_{k1}}{k_l} + C_{k0} \, k_{k1} + k_{l} + 2(k_+ \, \bar{c} + k_-)},
\end{aligned}
\end{equation}

which finally gives:
\begin{equation}
\begin{aligned}
\frac{\Delta c_{rms}}{\bar{c}} &>\, \left[ \frac{  I_{Ca} k_{k1} k_{dp}}{ (k_+ \, \bar{c} + k_- )+k_{dp}}  \right] ^{\frac{1}{2}}\frac{ 1}{ k_{d1} + \frac{I_{\text{Ca}} \,  \, k_{k1}}{k_l} + C_{k0} \, k_{k1} + k_{l} + 2(k_+ \, \bar{c} + k_-)}.
\end{aligned}
\label{append: lower bound}
\end{equation}
The mean concentrations, mean occupation probability, and the various parameters used in the above derivation are provided in Eq.~\ref{parameters_rms}.

\section{Cases When One Cytoplasmic Rate Constant \(\gg\) Other Cytoplasmic Rate Constants}
\label{appendE}

Rewrite the parameters in Eq.~\ref{parameters_rms} using the expression for 
$\bar{C}_{\text{Ca}} $ and $\bar{C}_{\text{K}}$:
\begin{equation}
\begin{aligned}
a &= k_{d1} + \frac{I_{\text{Ca}} \, \bar{p} \, k_{k1}}{k_{l}} + \frac{C_{k0} \, k_{d1} \, k_{l} \, k_{k1}}{k_{d1} \, k_{l} + I_{\text{Ca}} \, k_{k1} \, \bar{p}} + k_{l} \\
b &= \left( k_{d1} + \frac{I_{\text{Ca}} \, \bar{p} \, k_{k1}}{k_{l}} + \frac{C_{k0} \, k_{d1} \, k_{l} \, k_{k1}}{k_{d1} \, k_{l} + I_{\text{Ca}} \, k_{k1} \, \bar{p}} + k_{l} \right)^2 - 4 \left( k_{d1} k_{l} + I_{\text{Ca}} \, \bar{p} \, k_{k1} \right)\\
d &=k_{dp}\\
a^2-b &= 4 ( k_{d1} k_{l} + I_{\text{Ca}} \, \bar{p} \, k_{k1})\\
 a+d &=(k_{d1} + \frac{I_{\text{Ca}} \, \bar{p} \, k_{k1}}{k_{l}} + \frac{C_{k0} \, k_{d1} \, k_{l} \, k_{k1}}{k_{d1} \, k_{l} + I_{\text{Ca}} \, k_{k1} \, \bar{p}} + k_{l})+k_{dp}\\
a+2d &=(k_{d1} + \frac{I_{\text{Ca}} \, \bar{p} \, k_{k1}}{k_{l}} + \frac{C_{k0} \, k_{d1} \, k_{l} \, k_{k1}}{k_{d1} \, k_{l} + I_{\text{Ca}} \, k_{k1} \, \bar{p}} + k_{l})+2k_{dp}.
\end{aligned}
\label{parameters}
\end{equation}

\subsection{\( k_{k1} \gg \text{other cytoplasmic rate constants}\)}

When \( k_{k1} \gg \text{other cytoplasmic rate constants} \), the parameters described in Eq.~\ref{parameters} can be approximated as follows:
\begin{align}
a  & \approx \frac{I_{\text{Ca}} \, \bar{p} \, k_{k1}}{k_{l}} \nonumber \\
b  & \approx \left(\frac{I_{\text{Ca}} \, \bar{p} \, k_{k1}}{k_{l}}\right)^2-4 (I_{\text{Ca}} \, \bar{p} \, k_{k1})\nonumber\\
d &=k_{dp}\nonumber\\
a^2-b &\approx 4 I_{\text{Ca}} \, \bar{p} \, k_{k1} \approx 4ak_{l}\nonumber\\
a+d & \approx \frac{I_{\text{Ca}} \, \bar{p} \, k_{k1}}{k_{l}} \approx a \nonumber \\
 &a+2d \approx a\nonumber \\
&(a+2d)^2  \approx a^2+4ad \nonumber \\
(a+d)^2  & \approx a^2+2ad.
\end{align}
Using the above relations, Eq.~\ref{errorinconc3} simplifies to (utilizing the \(=\) sign in place of the \(\approx\) sign):
\begin{equation}
\begin{aligned}
\frac{\Delta c_{rms}}{\bar{c}} 
 &=  \, \bigg[ \frac{ d(a^2-b) \tau_c}{\bar{p} \, (1 - \bar{p})} \, \frac{4 (a + d) (1+ (a + d) \, \tau_c) + a \, [(a + 2d)^2-b] \, \tau_c^2}{a  \, [(a + 2d)^2-b] \, (1 + d \tau_c) \, [4 + 4a \tau_c + (a^2 - b) \tau_c^2]} \bigg]^{\frac{1}{2}}\\
 &=\, \bigg[ \frac{ d(a^2-b) \tau_c}{\bar{p} \, (1 - \bar{p})} \, \frac{4a(1+a\tau_c)+a(a^2+4ad-b)\tau_c^2}{a(a^2+4ad-b)(1+d\tau_c)(4+4a\tau_c+(a^2-b) \tau_c^2)}\bigg]^{\frac{1}{2}}\\
 &=\, \bigg[ \frac{ d(a^2-b) \tau_c}{\bar{p} \, (1 - \bar{p})} \, \frac{4(1+a\tau_c)+(a^2+4ad-b)\tau_c^2}{(a^2+4ad-b)(1+d\tau_c)(4+4a\tau_c+(a^2-b) \tau_c^2)}\bigg]^{\frac{1}{2}}\\
  &=\, \bigg[ \frac{ d(a^2-b) \tau_c}{\bar{p} \, (1 - \bar{p})} \, \frac{(4+4a\tau_c+(a^2-b) \tau_c^2)+4ad\tau_c^2}{(a^2+4ad-b)(1+d\tau_c)(4+4a\tau_c+(a^2-b) \tau_c^2)}\bigg]^{\frac{1}{2}}. \\
\end{aligned}
\label{largekk1}
\end{equation}
Substituting $(a^2-b)\approx 4ak_{l}$ values in above equation gives:
\begin{equation}
\begin{aligned}
\frac{\Delta c_{rms}}{\bar{c}}
&=\, \bigg[ \frac{ 4adk_{l} \tau_c}{\bar{p} \, (1 - \bar{p})} \, \frac{(4+4a\tau_c+4ak_{l} \tau_c^2)+4ad\tau_c^2}{(4ad+4ak_{l})(1+d\tau_c)(4+4a\tau_c+4ak_{l} \tau_c^2)}\bigg]^{\frac{1}{2}}\\
&=\, \bigg[ \frac{ 4adk_{l} \tau_c}{\bar{p} \, (1 - \bar{p})} \, \frac{(4a\tau_c+4ak_{l} \tau_c^2)+4ad\tau_c^2}{(4ad+4ak_{l})(1+d\tau_c)(4a\tau_c+4ak_{l} \tau_c^2)}\bigg]^{\frac{1}{2}}\\
&=\, \bigg[ \frac{ 4adk_{l} \tau_c}{\bar{p} \, (1 - \bar{p})} \, \frac{4a\tau_c(1+k_{l} \tau_c+d\tau_c)}{4a(d+k_{l})(1+d\tau_c)4a\tau_c(1+k_{l} \tau_c)}\bigg]^{\frac{1}{2}}\\
&=\, \bigg[ \frac{ dk_{l} \tau_c}{\bar{p} \, (1 - \bar{p})} \, \frac{(1+k_{l} \tau_c+d\tau_c)}{(d+k_{l})(1+d\tau_c)(
1+k_{l} \tau_c)}\bigg]^{\frac{1}{2}}.
\end{aligned}
\label{largekk2}
\end{equation}
Substituting $d=k_{dp}$ in above equation gives:
\begin{equation}
\begin{aligned}
\frac{\Delta c_{rms}}{\bar{c}} 
&=\, \bigg[ \frac{ k_{dp}k_{l} \tau_c}{\bar{p} \, (1 - \bar{p})} \, \frac{(1+k_{l} \tau_c+k_{dp}\tau_c)}{(k_{dp}+k_{l})(1+k_{dp}\tau_c)(
1+k_{l} \tau_c)}\bigg]^{\frac{1}{2}}.
\end{aligned}
\label{largekk3}
\end{equation}
Substituting $\bar{p}$ and $\tau_c$ from Eq.~\ref{parameters_rms} in above equation gives:
\begin{equation}
\begin{aligned}
\frac{\Delta c_{rms}}{\bar{c}} 
&=\, \bigg[ \frac{ k_{dp}k_{l} \tau_b}{\bar{p} \, } \, \frac{(1+k_{l} (1 - \bar{p})\tau_b+k_{dp}(1 - \bar{p})\tau_b)}{(k_{dp}+k_{l})(1+k_{dp}(1 - \bar{p})\tau_b)(
1+k_{l} (1 - \bar{p})\tau_b)}\bigg]^{\frac{1}{2}}\\
&=\, \bigg[ \frac{ k_{dp}k_{l} (k_+\bar{c} +k_-)}{k_-k_+\bar{c} \, } \, \frac{(1+\frac{k_{l}}{ (k_+ \bar{c} +k_-)}+\frac{k_{dp}}{ (k_+ \bar{c}+k_-)})}{(k_{dp}+k_{l})(1+\frac{k_{dp}}{ (k_+ \bar{c} +k_-)})(
1+\frac{k_{l}}{ (k_+ \bar{c} +k_-)})}\bigg]^{\frac{1}{2}}.
\end{aligned}
\label{largekk4}
\end{equation}

\subsection{\( k_{d1} \gg \text{other cytoplasmic rate constants}\)}

Under the condition that \( k_{d1} \gg \text{other cytoplasmic rate constants} \), the parameters described in Eq.~\ref{parameters} can be approximated as:
\begin{align}
a & \approx k_{d1}\nonumber\\
b & \approx \left(k_{d1}\right)^2-4 k_{d1}k_l\nonumber\\
d & =k_{dp}\nonumber\\
a^2-b & \approx 4 k_{d1}k_l \approx 4ak_l \nonumber\\
a+d & \approx k_{d1} \approx a\nonumber\\
 a+2d &\approx  a\\
(a+2d)^2 & \approx a^2+4ad \nonumber\\
(a+d)^2  & \approx a^2+2ad.
\end{align}
By using the above relations, Eq.~\ref{errorinconc3} reduces to (utilizing the \(=\) sign in place of the \(\approx\) sign):
\begin{equation}
\begin{aligned}
\frac{\Delta c_{rms}}{\bar{c}} = \bigg[ \frac{ dk_{l} \tau_c}{\bar{p} \, (1 - \bar{p})} \, \frac{(1+k_{l} \tau_c+d\tau_c)}{(d+k_{l})(1+d\tau_c)(1+k_{l} \tau_c)}\bigg]^{\frac{1}{2}} =\, \bigg[ \frac{ k_{dp}k_{l} \tau_c}{\bar{p} \, (1 - \bar{p})} \, \frac{(1+k_{l} \tau_c+k_{dp}\tau_c)}{(k_{dp}+k_{l})(1+k_{dp}\tau_c)(1+k_{l} \tau_c)}\bigg]^{\frac{1}{2}}.
\end{aligned}
\label{kd1larger1}
\end{equation}
Substituting $\bar{p}$ and $\tau_c$ from Eq.~\ref{parameters_rms} in above equation gives:
\begin{equation}
\begin{aligned}
\frac{\Delta c_{rms}}{\bar{c}} &= \bigg[ \frac{ k_{dp}k_{l} \tau_b}{\bar{p} \, } \, \frac{(1+k_{l} (1 - \bar{p})\tau_b+k_{dp}(1 - \bar{p})\tau_b)}{(k_{dp}+k_{l})(1+k_{dp}(1 - \bar{p})\tau_b)(1+k_{l} (1 - \bar{p})\tau_b)}\bigg]^{\frac{1}{2}} \\
&= \bigg[ \frac{ k_{dp}k_{l} (k_+\bar{c} +k_-)}{k_-k_+\bar{c} \, } \, \frac{(1+\frac{k_{l}}{ (k_+\bar{c} +k_-)}+\frac{k_{dp}}{ (k_+\bar{c} +k_-)})}{(k_{dp}+k_{l})(1+\frac{k_{dp}}{ (k_+\bar{c} +k_-)})(
1+\frac{k_{l}}{ (k_+\bar{c} +k_-)})}\bigg]^{\frac{1}{2}}.
\end{aligned}
\label{kd1larger2}
\end{equation}

\subsection{\( k_{l} \gg \text{other cytoplasmic rate constants}\)}

If \( k_{l} \gg \text{other cytoplasmic rate constants}\), the parameters detailed in Eq.~\ref{parameters} can be approximated as follows:
\begin{align}
a &\approx k_{l}\nonumber\\
b &\approx \left(k_{l}\right)^2-4 k_{d1}k_l\nonumber\\
d &=k_{dp}\nonumber\\
a^2-b &\approx 4 k_{d1}k_l \approx 4ak_{d1}\nonumber\\
a+d   &\approx k_{l} =a\\
 a+2d &\approx a
\end{align}
Utilizing the above relations, Eq.~\ref{errorinconc3} can be approximated as (using the \(=\) sign in place of the \(\approx\) sign):
\begin{equation}
\begin{aligned}
\frac{\Delta c_{rms}}{\bar{c}} 
&=\, \bigg[ \frac{ d(a^2-b) \tau_c}{\bar{p} \, (1 - \bar{p})} \, \frac{(4+4a\tau_c+(a^2-b) \tau_c^2)+4ad\tau_c^2}{(a^2+4ad-b)(1+d\tau_c)(4+4a\tau_c+(a^2-b) \tau_c^2)}\bigg]^{\frac{1}{2}}\\
&=\, \bigg[ \frac{ d(a^2-b) \tau_c}{\bar{p} \, (1 - \bar{p})} \, \frac{(4a\tau_c+(a^2-b) \tau_c^2)+4ad\tau_c^2}{(a^2+4ad-b)(1+d\tau_c)(4a\tau_c+(a^2-b) \tau_c^2)}\bigg]^{\frac{1}{2}} \\
&=\, \bigg[ \frac{ 4adk_{d1} \tau_c}{\bar{p} \, (1 - \bar{p})} \, \frac{(4a\tau_c+4ak_{d1}\tau_c^2)+4ad\tau_c^2}{(4ak_{d1}+4ad)(1+d\tau_c)(4a\tau_c+4ak_{d1} \tau_c^2)}\bigg]^{\frac{1}{2}}\\
&=\, \bigg[ \frac{ dk_{d1} \tau_c}{\bar{p} \, (1 - \bar{p})} \, \frac{(1+k_{d1}\tau_c+d\tau_c)}{(k_{d1}+d)(1+d\tau_c)(1+k_{d1} \tau_c)}\bigg]^{\frac{1}{2}}.
\end{aligned}
\label{append:largekl}
\end{equation}
Substituting $d=k_{dp}$ in above equation gives:
\begin{equation}
\begin{aligned}
\frac{\Delta c_{rms}}{\bar{c}} 
&=\, \bigg[ \frac{ k_{dp}k_{d1} \tau_c}{\bar{p} \, (1 - \bar{p})} \, \frac{(1+k_{d1}\tau_c+k_{dp}\tau_c)}{(k_{d1}+k_{dp})(1+k_{dp}\tau_c)(1+k_{d1} \tau_c)}\bigg]^{\frac{1}{2}}.
\end{aligned}
\label{largekl2}
\end{equation}
Substituting $\bar{p}$ and $\tau_c$ from Eq.~\ref{parameters_rms} in above equation gives:
\begin{equation}
\begin{aligned}
\frac{\Delta c_{rms}}{\bar{c}} 
&=\, \bigg[ \frac{ k_{dp}k_{d1} \tau_b}{\bar{p} \, } \, \frac{(1+k_{d1} (1 - \bar{p})\tau_b+k_{dp}(1 - \bar{p})\tau_b)}{(k_{dp}+k_{d1})(1+k_{dp}(1 - \bar{p})\tau_b)(
1+k_{d1} (1 - \bar{p})\tau_b)}\bigg]^{\frac{1}{2}}\\
&=\, \bigg[ \frac{ k_{dp}k_{d1} (k_+\bar{c} +k_-)}{k_-k_+\bar{c} \, } \, \frac{(1+\frac{k_{d1}}{ (k_+\bar{c} +k_-)}+\frac{k_{dp}}{ (k_+\bar{c} +k_-)})}{(k_{dp}+k_{l})(1+\frac{k_{dp}}{ (k_+\bar{c} +k_-)})(
1+\frac{k_{d1}}{ (k_+\bar{c} +k_-)})}\bigg]^{\frac{1}{2}}.
\end{aligned}
\label{largekl3}
\end{equation}

\begin{figure}[htpb]
    \centering
    \begin{minipage}[b]{0.21\textwidth}
        \centering
      \raggedright\textbf{(a)}   \includegraphics[width=\textwidth]{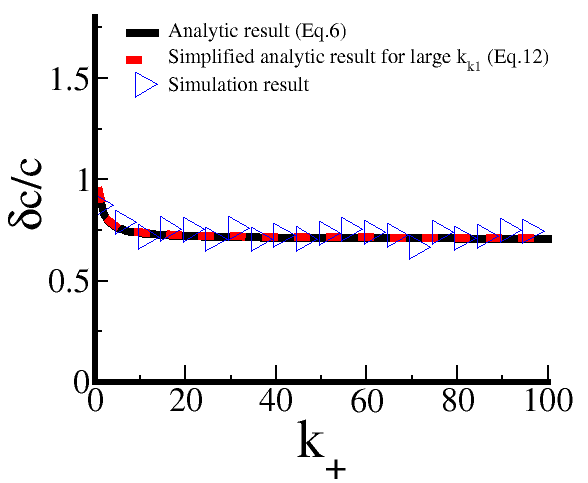}
        \label{fig:image-a}
    \end{minipage}
    \hspace{0.01\textwidth}
    \begin{minipage}[b]{0.21\textwidth}
        \centering
   \raggedright\textbf{(b)}      \includegraphics[width=\textwidth]{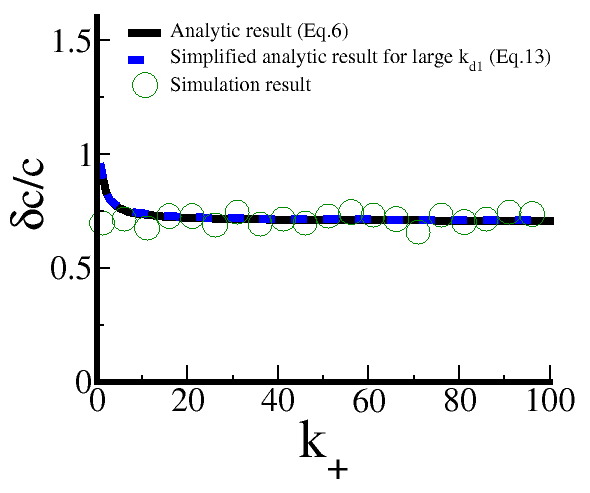}
        \label{fig:image-b}
    \end{minipage}
    \hspace{0.01\textwidth}
    \begin{minipage}[b]{0.21\textwidth}
        \centering
     \raggedright\textbf{(c)}    \includegraphics[width=\textwidth]{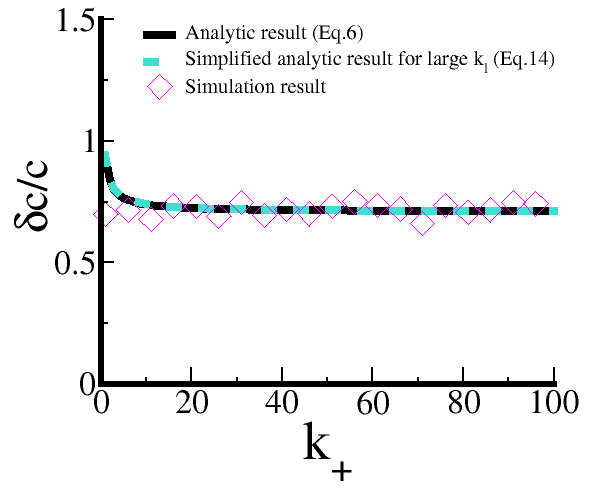}
        \label{fig:image-c}
    \end{minipage}
    \caption{  Error ($\delta c / c$) vs. ligand attachment rate ($k_+$) for different cytoplasmic rate constants. The plot demonstrates how error varies with the ligand attachment rate ($k_+$) in three scenarios where specific cytoplasmic rate constants are set to 50( \(k_{k1}\) in units of \(\text{M}^{-1}\text{s}^{-1}\); $I_{Ca}$ in units of $Ms^{-1}$,  and the others in \(\text{s}^{-1}\) ), which is significantly larger than the other cytoplasmic rate constants . All cytoplasmic rate constants ($k_{k1},k_{d1}, k_l, k_p, k_{dp}$) are set to 1 (units specified as above), except for the rate constant being varied in each scenario. In scenario (a), $k_{k1} = 50M^{-1}s^{-1}$; in scenario (b), $k_{d1} = 50s^{-1}$; and in scenario (c), $k_l = 50s^{-1}$.  In all cases,ligand detachment rate  $k_- = 1s^{-1}$, calcium influx rate $ I_{Ca}= 1Ms^{-1}$,with initial kinase concentration set to 1M.}
    \label{limitingcase_kplusplot}
\end{figure}

\begin{figure}[htpb]
    \centering
    \begin{minipage}[b]{0.21\textwidth}
        \centering
        \includegraphics[width=\textwidth]{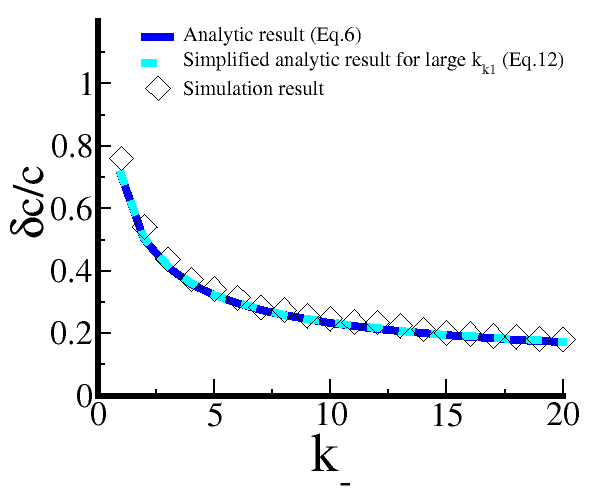}
        \label{fig:image-a}
    \end{minipage}
    \hspace{0.05\textwidth}
    \begin{minipage}[b]{0.21\textwidth}
        \centering
        \includegraphics[width=\textwidth]{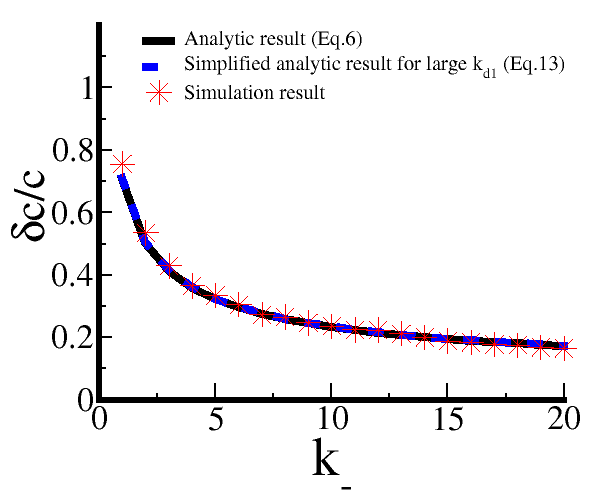}
        \label{fig:image-b}
    \end{minipage}
 \hspace{0.05\textwidth}
    \begin{minipage}[b]{0.21\textwidth}
        \centering
        \includegraphics[width=\textwidth]{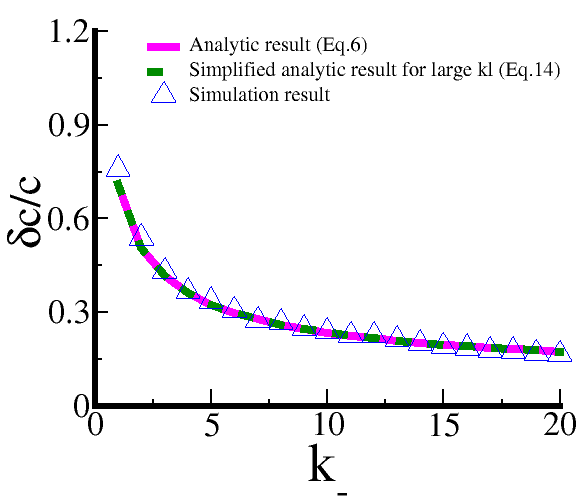}
        \label{fig:image-c}
    \end{minipage}
 \caption{  Error ($\delta c / c$) vs. ligand detachment rate ($k_-$) for different cytoplasmic rate constants. The plot demonstrates how error varies with the ligand detachment rate ($k_-$) in three scenarios where specific cytoplasmic rate constants are set to 50( \(k_{k1}\) in units of \(\text{M}^{-1}\text{s}^{-1}\); $I_{Ca}$ in units of $Ms^{-1}$,  and the others in \(\text{s}^{-1}\) ), which is significantly larger than the other cytoplasmic rate constants . All cytoplasmic rate constants ($k_{k1},k_{d1}, k_l, k_p, k_{dp}$) are set to 1(units specified as above), except for the rate constant being varied in each scenario. In scenario (a), $k_{k1} = 50M^{-1}s^{-1}$; in scenario (b), $k_{d1} = 50s^{-1}$; and in scenario (c), $k_l = 50s^{-1}$.  In all cases,ligand attachment rate  $k_+ = 100M^{-1}s^{-1}$, calcium influx rate $ I_{Ca}= 1Ms^{-1}$, with initial kinase concentration also set to 1M. Note: Eqs. E6, E9, and E13 are same as Eqs.12, 13 and 14 respectively.}
    \label{limitingcase_kminusplot}
\end{figure}

Figs. \ref{limitingcase_kplusplot} and \ref{limitingcase_kminusplot} illustrate the steady-state error ($\delta c / c$) in a cellular signaling system as influenced by variations in ligand attachment rate ($k_+$) and detachment rate ($k_-$) under specific conditions where certain cytoplasmic rate constants are significantly larger than others. Fig. {\ref{limitingcase_kplusplot}} shows the error as a function of $k_+$ when the calcium binding rate to calmodulin ($k_{k1}$), the calcium unbinding rate from the kinase-calcium complex ($k_{d1}$), or the calcium loss rate ($k_l$) is much larger than other cytoplasmic rates (Eq. 12, Eq. 13 and Eq. 14). Fig. \ref{limitingcase_kminusplot} examines the error as a function of $k_-$ under similar conditions where the same rate constants are dominant. We compared the analytic result for the RMS error (Eq. 6) with the simplified analytic results for RMS error (Eqs. E6, E9, and E13), as well as with the simulation results, and found strong agreement across all three approaches.

\section{Mean Calculation Once the Cell Reaches Steady State}
\label{appendF}

The mean concentrations at steady state can be obtained by taking the expectation value of both sides of Eq.~2-5. At steady state, the mean concentration remains constant and does not vary with time. Therefore, setting the time derivative to zero yields:
\begin{equation}
\begin{aligned}
I_{Ca} \, \langle p \rangle - k_l \langle C_{Ca} \rangle - k_{k1} \langle C_K \, C_{Ca} \rangle + k_{d1} \langle C_{K-Ca} \rangle &=0 \\
 k_{d1} \langle C_{K-Ca} \rangle - k_{k1} \langle C_{K} \, C_{Ca} \rangle &=0 \\
 k_{k1} \langle C_{K} \, C_{Ca} \rangle - k_{d1} \langle C_{K-Ca} \rangle  &=0 \\
k_p \langle C_{K-Ca} \rangle - k_{dp} \langle C_{Pr} \rangle  &=0.
\end{aligned}
\label{mean_concentration_nonlinear}
\end{equation}
Using Eq.~\ref{CkCca_expectation}, along with \( \langle C \rangle = \bar{C} \), and replacing the approximation sign (\(\approx\)) with the equality sign (\(=\)) for further calculations, the above expression becomes:
\begin{equation}
\begin{aligned}
I_{Ca} \, \bar{p} - k_l \bar{C}_{Ca} - k_{k1} \bar{C}_K \bar{C}_{Ca} + k_{d1} \bar{C}_{K-Ca} &= 0 \\
k_{d1} \bar{C}_{K-Ca} - k_{k1} \bar{C}_K \bar{C}_{Ca} &= 0 \\
k_{k1} \bar{C}_K \bar{C}_{Ca} - k_{d1} \bar{C}_{K-Ca} &= 0 \\
k_p \bar{C}_{K-Ca} - k_{dp} \bar{C}_{Pr} &= 0.
\end{aligned}
\label{mean_concentration_linear}
\end{equation}
Adding Eq.~3 and Eq.~4 gives:
\begin{equation}
\begin{split}
\frac{d}{dt}(C_{K}(t)+C_{K-Ca}(t)) = 0  \Longrightarrow C_{K}(t)+C_{K-Ca}(t) = C_{K0},
\end{split} 
\end{equation}
where \( C_{K0} \) denotes the initial concentration of calmodulin. The initial concentration of \( C_{K-Ca} \) is assumed to be zero. Taking expectation value on both side of above equation gives:
\begin{equation}
\begin{split}
\bar{C}_K +\bar{C}_{K-Ca} = C_{K0}.
\end{split} 
\label{CkCk-camean}
\end{equation}
Solving Eqs. \ref{mean_concentration_linear} and \ref{CkCk-camean} for the four unknowns \( \bar{C}_{\text{Ca}} \), \( \bar{C}_{\text{K}} \), \( \bar{C}_{\text{K-Ca}} \), and \( \bar{C}_{\text{Pr}} \) yields:
\begin{equation}
\begin{split}
\bar{C}_{\text{Ca}} &= \frac{I_{\text{Ca}} \, \bar{p} }{k_l} \\
\bar{C}_{\text{K}} &= \frac{C_{k0} \, k_{d1} \, k_l}{k_{d1} \, k_l + I_{\text{Ca}} \, k_{k1} \, \bar{p}} \\
\bar{C}_{\text{K-Ca}} &= \frac{C_{k0} \, I_{\text{Ca}} \, k_{k1} \, \bar{p}}{k_{d1} \, k_l + I_{\text{Ca}} \, k_{k1} \, \bar{p}} \\
\bar{C}_{\text{Pr}}&= \frac{C_{K0} \, I_{\text{Ca}} \, k_{k1} \, k_{p} \, \bar{p}}{k_{dp} \, (k_{d1} \, k_l + I_{\text{Ca}} \,  k_{k1} \, \bar{p})}.
\end{split} 
\label{mean_concentrations}
\end{equation}
The value \(\bar{p}\) in the above equation can be determined by taking the expectation value of both sides of the equation \(\frac{dp(t)}{dt} = k_+ c \, (1 - p(t)) - k_- p(t)\), which describes the time evolution of the receptor's occupation probability \cite{Bialek2005}:
\begin{equation}
\frac{d \langle p(t) \rangle}{dt} = k_+ \, c \, (1 - \langle p(t) \rangle) - k_- \, \langle p(t) \rangle.
\end{equation}
At steady state, setting \(\frac{d \langle p(t) \rangle}{dt} = 0\), \(\langle p(t) \rangle = \bar{p}\), and \(c = \bar{c}\) (assuming the cell is under a uniform ligand concentration) gives:

\begin{equation}
\bar{p} = \frac{k_+ \bar{c}}{k_+ \bar{c} + k_-}.
\label{barp}
\end{equation}

\section{Relationship Between RMS Error in \(C_{Pr}\) and RMS Error in \(c\)}
\label{appendG}

To calculate the RMS error in extracellular ligand concentration \( c \) through phosphorylation readout \( C_{Pr} \), a relationship between \( c \) and \( C_{Pr} \) is defined as follows:
\begin{equation}
\begin{aligned}
\frac{\delta c}{c} &= \frac{\delta C_{\text{Pr}}}{c \, \frac{\delta C_{\text{Pr}}}{\delta c}}. 
\end{aligned}
\end{equation}
This relationship tells us how changes in \( C_{\text{Pr}} \) affect \( c \). Here \( \frac{\delta c}{c} \) represents the relative change in \( c \). \( \delta C_{\text{Pr}} \) represents the change in \( C_{\text{Pr}} \), which could be a small or infinitesimal change in \( C_{\text{Pr}} \). \( \frac{\delta C_{\text{Pr}}}{\delta c} \) represents the rate of change of \( C_{\text{Pr}} \) with respect to \( c \), often denoted as the derivative $\frac{dC_{Pr}}{dc}$. Squaring both side of above equation gives: 
\begin{equation}
\begin{aligned}
\frac{(\delta c)^2}{c^2} &= \frac{(\delta C_{\text{Pr}})^2}{c^2 \, \left(\frac{d C_{\text{Pr}}}{d c} \right)^2 }. 
\end{aligned}
\end{equation}
above equation can be written as:
\begin{equation}
\begin{aligned}
\frac{\sigma_c^2}{\bar{c}^2} &= \frac{\sigma_{C_{\text{Pr}}}^2}{\bar{c}^2 \, \left(\frac{d \bar{C}_{\text{Pr}}}{d \bar{c}} \right)^2 }. 
\end{aligned}
\label{rmserror_c}
\end{equation} 
Letting \(B_1 = C_{K0} \, I_{Ca} \, k_{k1} \, k_{p}\), \(B_2 = k_{dp} \, k_{d1} \, k_{l}\), and \(B_3 = k_{dp} \, I_{Ca} \, k_{k1}\), Eq.~\ref{mean_concentrations} gives:
\begin{equation}
\begin{aligned}
\bar{C}_{\text{Pr}} = \frac{B_1 \, \bar{p}}{B_2 + B_3 \, \bar{p}} = \frac{1}{\frac{B_3}{B_1} + \frac{B_2}{B_1 \, \bar{p}}} = \frac{1}{\frac{B_3}{B_1} + \frac{B_2}{B_1} \, \left( 1 + \frac{k_-}{k_+ \, \bar{c}} \right)}.
\end{aligned}
\end{equation}
Differentiating with respect to $\bar{c}$ gives:
\begin{equation}
\begin{aligned}
\frac{d \bar{C}_{\text{Pr}}}{d \bar{c}} &= \frac{-\frac{d}{d \bar{c}} \left[ \frac{B_3}{B_1} + \frac{B_2}{B_1} \, \left( 1 + \frac{k_-}{k_+ \, \bar{c}} \right)\right]}{\left[\frac{B_3}{B_1} + \frac{B_2}{B_1} \, \left( 1 + \frac{k_-}{k_+ \, \bar{c}} \right)\right]^2} = \frac{\frac{B_2}{B_1} \, \left( \frac{k_-}{k_+ \, \bar{c}^2} \right)}{\left[\frac{B_3}{B_1} + \frac{B_2}{B_1} \, \left( 1 + \frac{k_-}{k_+ \, \bar{c}} \right)\right]^2} = \frac{B_2}{B_1} \, \left( \frac{k_-}{k_+ \, \bar{c}^2} \right) \, \bar{C}_{\text{Pr}}^2.
\end{aligned}
\end{equation}
Substituting above relation back into \ref{rmserror_c} yields:
\begin{equation}
\begin{aligned}
\frac{\sigma_c^2}{\bar{c}^2} = \left( \frac{k_+ \, \bar{c}}{k_-} \right)^2 \, \left(\frac{B_1}{B_2} \right)^2 \, \frac{\sigma_{C_{\text{Pr}}}^2}{\bar{C}_{\text{Pr}}^4} = \left( \frac{\bar{p}}{1-\bar{p}} \right)^2 \, \left(\frac{B_1}{B_2} \right)^2 \, \frac{\sigma_{C_{\text{Pr}}}^2}{\bar{C}_{\text{Pr}}^4},
\end{aligned}
\end{equation}
which gives the RMS error as follows:
\begin{equation}
\begin{aligned}
\frac{\Delta c_{rms}}{\bar{c}} = \frac{\sigma_c}{\bar{c}} &= \left( \frac{\bar{p}}{1-\bar{p}} \right) \, \left(\frac{B_1}{B_2} \right) \, \frac{\sigma_{C_{\text{Pr}}}}{\bar{C}_{\text{Pr}}^2}.
\end{aligned}
\end{equation}
Substituting the predefined notations \( B_1 = C_{K0} \, I_{\text{Ca}} \, k_{k1} \, k_{p} \), \( B_2 = k_{dp} \, k_{d1} \, k_l \), and \( \bar{C}_{\text{Pr}} \) (from Eq.~\ref{mean_concentrations}) into the above equation, and then simplifying, gives:
\begin{equation}
\begin{aligned}
\frac{\Delta c_{rms}}{\bar{c}} = \frac{\sigma_c}{\bar{c}} &= \frac{{k_{dp} \, (k_{d1} \, k_l + I_{\text{Ca}} \, k_{k1} \, \bar{p})^2}}{{C_{K0} \, I_{\text{Ca}} \, k_{d1} \, k_{k1} \, k_l \, k_p \, \bar{p} \, (1 - \bar{p})}} \, \sigma_{C_{\text{Pr}}}.
\end{aligned}
\end{equation}
Above equation along with \( \bar{C}_{\text{Pr}} \) (from Eq.~\ref{mean_concentrations}) gives:
\begin{equation}
\begin{aligned}
\frac{(\Delta C_{Pr})_{rms}}{\bar{C}_{Pr}} = \frac{\sigma_{C_{Pr}}}{\bar{C}_{Pr}} = \frac{C_{K0} \, I_{Ca} k_{d1} \, k_{k1} \, k_l \, k_p \, \bar{p} \, (1 - \bar{p})}{k_{dp} \, (k_{d1} \, k_l + I_{Ca} \, k_{k1} \, \bar{p})^2} \, \frac{k_{dp} \, (k_{d1} \, k_l + I_{Ca} \,  k_{k1} \, \bar{p})}{C_{K0} \, I_{Ca} \, k_{k1} \, k_{p} \, \bar{p}}  \, \frac{\Delta c_{rms}}{\bar{c}},
\end{aligned}
\end{equation}
which simplifies to:
\begin{equation}
\begin{aligned}
\frac{(\Delta C_{Pr})_{rms}}{\bar{C}_{Pr}} = \bigg( \frac{k_{d1} \, k_{l} \, (1-\bar{p})}{k_{d1} \, k_{l}  + I_{Ca} \,  k_{k1} \, \bar{p}} \bigg) \, \frac{\Delta c_{rms}}{\bar{c}}.
\end{aligned}
\label{Cpr_error1}
\end{equation}
Substituting $\bar{p}$ from Eq.~\ref{parameters_rms} in above equation gives:
\begin{equation}
\begin{aligned}
\frac{(\Delta C_{Pr})_{rms}}{\bar{C}_{Pr}} = \bigg( \frac{k_{d1} \, k_l \, k_-}{k_{d1} \, k_{l} \, (k_+ \bar{c} + k_-) + I_{Ca} \, k_{k1} \, k_+ \bar{c}} \bigg) \, \frac{\Delta c_{rms}}{\bar{c}}.
\end{aligned}
\label{Cpr_error2}
\end{equation}

\textbf{The expression for $\frac{(\Delta C_{Ca})_{\mathrm{rms}}}{\bar{C}_{Ca}}$ is derived as:} The relationship between the error in $c$ and the error in $C_{Ca}$, analogous to Eq.~\ref{rmserror_c}, can be expressed as:
\begin{equation}
\begin{aligned}
\frac{\sigma_c^2}{\bar{c}^2} &= \frac{\sigma_{C_{Ca}}^2}{\bar{c}^2 \, \left(\frac{d \bar{C}_{Ca}}{d \bar{c}} \right)^2 }. 
\end{aligned}
\label{rmserror_c_ca}
\end{equation}
Eq.~\ref{parameters_rms} gives:
\begin{equation}
\begin{aligned}
\bar{C}_{\text{Ca}} = \frac{I_{\text{Ca}} \, \bar{p} }{k_l} = \frac{I_{\text{Ca}}}{k_l} \, \frac{k_+ \, \bar{c}}{k_+ \, \bar{c} + k_-}.
\end{aligned}
\end{equation}
Differentiating with respect to $\bar{c}$ gives:
\begin{equation}
\begin{aligned}
\frac{d \bar{C}_{\text{Ca}}}{d \bar{c}} &= \frac{I_{\text{Ca}}}{k_l} \, \frac{(k_+ \, \bar{c} + k_-) \, k_+ - k_+ \, \bar{c} \, k_+}{(k_+ \, \bar{c} + k_-)^2} = \frac{I_{\text{Ca}}}{k_l} \, \frac{k_+ \, k_-}{(k_+ \, \bar{c} + k_-)^2} = \frac{I_{\text{Ca}}}{k_l \, \bar{c}} \, \frac{k_+ \bar{c} \, k_-}{(k_+ \, \bar{c} + k_-)^2},
\end{aligned}
\end{equation}
which can be written in terms of $\bar{p}$ and $\bar{C}_{\text{Ca}}$ using Eq.~\ref{parameters_rms} as follows:
\begin{equation}
\begin{aligned}
\frac{d \bar{C}_{\text{Ca}}}{d \bar{c}} = \frac{I_{\text{Ca}}}{k_l \, \bar{c}} \, \bar{p} \, (1-\bar{p}) = \frac{\bar{C}_{\text{Ca}} \, (1-\bar{p})}{\bar{c}}.
\end{aligned}
\end{equation}
Substituting above equation back into Eq. \ref{rmserror_c_ca} gives;
\begin{equation}
\begin{aligned}
\frac{(\Delta C_{Ca})_{rms}}{\bar{C}_{Ca}} &= (1-\bar{p}) \, \frac{\Delta c_{rms}}{\bar{c}}.
\end{aligned}
\label{Cca_error}
\end{equation}

\textbf{The expression for $\frac{(\Delta C_{K})_{\mathrm{rms}}}{\bar{C}_{K}}$ is derived as:} The relationship between the error in $c$ and the error in $C_{K}$, similar to Eq.~\ref{rmserror_c}, can be written as:
\begin{equation}
\begin{aligned}
\frac{\sigma_c^2}{\bar{c}^2} &= \frac{\sigma_{C_{K}}^2}{\bar{c}^2 \, \left(\frac{d \bar{C}_{K}}{d \bar{c}} \right)^2 }. 
\end{aligned}
\label{rmserror_c_k}
\end{equation}
Differentiating $\bar{C}_{\text{K}}$ (from Eq.~\ref{parameters_rms}) with respect to $\bar{c}$ gives:
\begin{equation}
\begin{aligned}
\frac{d \bar{C}_{\text{K}}}{d \bar{c}} &= \frac{-C_{K0} \, k_{d1} \, k_l \, I_{\text{Ca}} \, k_{k1} \, \frac{d\bar{p}}{d\bar{c}} }{(k_{d1} \, k_l + I_{\text{Ca}} \, k_{k1} \, \bar{p})^2} = - \bar{C}_{\text{K}} \, \frac{I_{\text{Ca}} \, k_{k1} \, \frac{d\bar{p}}{d\bar{c}} }{k_{d1} \, k_l + I_{\text{Ca}} \, k_{k1} \, \bar{p}} = - \bar{C}_{\text{K}} \, \frac{I_{\text{Ca}} \, k_{k1}}{k_{d1} \, k_l + I_{\text{K}} \, k_{k1} \, \bar{p}} \, \frac{\bar{p} \, (1-\bar{p})}{\bar{c}}.
\end{aligned}
\end{equation}
Substituting above equation back into Eq. \ref{rmserror_c_k} gives;
\begin{equation}
\begin{aligned}
\frac{(\Delta C_{K})_{rms}}{\bar{C}_{K}} &= \frac{I_{Ca} \, k_{k1} \, \bar{p} \, (1-\bar{p})}{k_{d1} \, k_{l}  + I_{Ca} \,  k_{k1} \, \bar{p}} \, \frac{\Delta c_{rms}}{\bar{c}}.
\end{aligned}
\label{Ck_error}
\end{equation}

\section{Constraints to Hold Linearization Approximation}
\label{appendH}

Our analytical results in the main text are valid under the assumptions stated in Eq.~\ref{approx}, which for an ensemble of many cells can be expressed as:
\begin{equation}
\begin{aligned}
  \bar{C}_{Ca} \, \sigma_{C_K} + \bar{C}_{K} \, \sigma_{C_{Ca}} \gg \sigma_{C_{Ca}} \, \sigma_{C_K},
\end{aligned}
\end{equation}
where $\sigma_{C_{Ca}}$ and $\sigma_{C_{K}}$ represent the standard deviations of $C_{Ca}$ and $C_{K}$ across the ensemble of cells. Above inequality can be written as:
\begin{equation}
\begin{aligned}
\frac{\bar{C}_{Ca}}{\sigma_{C_{Ca}}} + \frac{\bar{C}_{K}}{\sigma_{C_{K}}} \gg 1 
\Longrightarrow \bigg[\frac{(\Delta C_{Ca})_{rms}}{\bar{C}_{Ca}} \bigg]^{-1} + \bigg[\frac{(\Delta C_{K})_{rms}}{\bar{C}_{K}} \bigg]^{-1} \gg 1.
\end{aligned}
\label{approx_ensemble}
\end{equation}
Utilizing Eqs.~\ref{Cca_error} and \ref{Ck_error} in the above equation yields:
\begin{equation}
\begin{aligned}
\bigg( \frac{1}{1-\bar{p}} + \frac{k_{d1} \, k_{l}  + I_{Ca} \,  k_{k1} \, \bar{p}}{I_{Ca} \, k_{k1} \, \bar{p} \, (1-\bar{p})} \bigg) \, \bigg( \frac{\Delta c_{rms}}{\bar{c}} \bigg)^{-1} \gg 1 
\Longrightarrow \bigg( \frac{I_{Ca} \, k_{k1} \, \bar{p}  + k_{d1} \, k_{l}  + I_{Ca} \,  k_{k1} \, \bar{p}}{I_{Ca} \, k_{k1} \, \bar{p} \, (1-\bar{p})} \bigg) \, \bigg( \frac{\Delta c_{rms}}{\bar{c}} \bigg)^{-1} \gg 1,
\end{aligned}
\end{equation}
which can be written as:
\begin{equation}
\begin{aligned}
\frac{I_{Ca} \, k_{k1} \, \bar{p} \, (1-\bar{p})}{2 \, I_{Ca} \, k_{k1} \, \bar{p}  + k_{d1} \, k_{l}} \, \frac{\Delta c_{rms}}{\bar{c}} & \ll 1,
\end{aligned}
\label{constraint_general}
\end{equation}
where \(\frac{\Delta c_{\text{rms}}}{\bar{c}}\) is defined by Eq.~\ref{errorinconc3}. Above inequality represents the constraint under which our general result for the RMS error in sensing ligand concentration (Eq. 6) is valid.

\subsection{$k_{k1} \gg \text{other cytoplasmic rate constants }$}

For \(k_{k1} \gg \text{other cytoplasmic rate constants}\), the prefactor in Eq.~\ref{constraint_general} simplifies to \(\frac{I_{Ca} \, k_{k1} \, \bar{p} \, (1-\bar{p})}{2 \, I_{Ca} \, k_{k1} \, \bar{p} + k_{d1} \, k_{l}} \approx \frac{(1 - \bar{p})}{2}\). Substituting \(\frac{\Delta c_{\text{rms}}}{\bar{c}}\) from Eq.~\ref{largekk3} into Eq.~\ref{constraint_general} further simplifies it to:
\begin{equation}
\begin{aligned}
\frac{(1 - \bar{p})}{2} \, \left[ \frac{k_{dp} k_{l} \tau_c}{\bar{p} \, (1 - \bar{p})} \, \frac{1 + k_{l} \tau_c + k_{dp} \tau_c}{(k_{dp} + k_{l})(1 + k_{dp} \tau_c)  (1 + k_{l} \tau_c)} \right]^{\frac{1}{2}} & \ll 1.
\end{aligned}
\label{constraint_kk1_1}
\end{equation}
This inequality is the constraint for the validity of Eq. 12. Above inequality can be simplified as follows:
\begin{equation}
\begin{aligned}
\frac{(1 - \bar{p}) \,  k_{dp} k_{l} \tau_c }{4\bar{p}} \, \frac{1 + k_{l} \tau_c + k_{dp} \tau_c}{(k_{dp} + k_{l})(1 + k_{dp} \tau_c)  (1 + k_{l} \tau_c)}  & \ll  1 \Longrightarrow \frac{(1 - \bar{p}) \,  k_{dp} k_{l} \tau_c }{4\bar{p} \, (k_{dp} + k_{l})} \, \frac{1 + k_{l} \tau_c + k_{dp} \tau_c}{1 + k_{dp} \tau_c+ k_{l} \tau_c + k_{dp} k_{l} \tau_c}  & \ll  1. 
\end{aligned}
\label{constraint_kk1_2}
\end{equation}
It is important to note that the second fraction in the inequality above is $
\frac{1 + k_l \tau_c + k_{dp} \tau_c}{1 + k_l \tau_c + k_{dp} \tau_c + k_{dp} k_l \tau_c} \leq 1$, which allows us to simplify the inequality by focusing on the first fraction, as the second fraction will not violate the inequality. Thus, we need only ensure that the following constraint holds:
\begin{equation}
\begin{aligned}
\frac{(1 - \bar{p}) \, \tau_c}{4\bar{p}} \, \frac{ \,  k_{dp} k_{l}}{k_{dp} + k_{l}}  & \ll  1.
\end{aligned}
\label{constraint_kk1_3}
\end{equation}
Substituting $\bar{p}$ and $\tau_c$ from Eq.~\ref{parameters_rms} in above equation gives:
\begin{equation}
\begin{aligned}
\frac{k_-}{k_+ \, \bar{c} \, (k_+ \, \bar{c} + k_-)} \, \frac{ \,  k_{dp} k_{l}}{k_{dp} + k_{l}}  & \ll  1.
\end{aligned}
\label{constraint_kk1_4}
\end{equation}
This inequality suggests that improved linearization is achieved when \(k_{dp}\), \(k_{l}\), and \(k_{-}\) have lower values, while \(k_{+} \, \bar{c}\) has larger value.

\subsection{$k_{d1} \gg \text{other cytoplasmic rate constants }$}

When \(k_{d1} \gg \text{other cytoplasmic rate constants}\), the prefactor in Eq.~\ref{constraint_general} reduces to $\frac{I_{Ca} \, k_{k1} \, \bar{p} \, (1-\bar{p})}{2 \, I_{Ca} \, k_{k1} \, \bar{p} + k_{d1} \, k_{l}} \approx \frac{I_{Ca} \, k_{k1} \, \bar{p} \, (1-\bar{p})}{k_{d1} \, k_{l}}.$ Substituting \(\frac{\Delta c_{\text{rms}}}{\bar{c}}\) from Eq.~\ref{kd1larger1} into Eq.~\ref{constraint_general} leads to further simplification:
\begin{equation}
\begin{aligned}
\frac{I_{Ca} \, k_{k1} \, \bar{p} \, (1-\bar{p})}{k_{d1} \, k_{l}}  \, \left[ \frac{k_{dp} k_{l} \tau_c}{\bar{p} \, (1 - \bar{p})} \, \frac{1 + k_{l} \tau_c + k_{dp} \tau_c}{(k_{dp} + k_{l})(1 + k_{dp} \tau_c)  (1 + k_{l} \tau_c)} \right]^{\frac{1}{2}} & \ll 1.
\end{aligned}
\label{constraint_kd1_1}
\end{equation}
This inequality serves as a constraint for the validity of Eq. 13. It can be further simplified as follows:
\begin{equation}
\begin{aligned}
\frac{\bar{p} \, (1-\bar{p}) \, \tau_c \, I_{Ca}^2 \, k_{k1}^2 \, k_{dp}}{k_{d1}^2 \, k_{l}}  \, \frac{1 + k_{l} \tau_c + k_{dp} \tau_c}{(k_{dp} + k_{l})(1 + k_{dp} \tau_c)  (1 + k_{l} \tau_c)}  \ll  1 \Longrightarrow \frac{\bar{p} \, (1-\bar{p}) \, \tau_c \, I_{Ca}^2 \, k_{k1}^2 \, k_{dp}}{k_{d1}^2 \, k_{l} \, (k_{dp} + k_l)}  \, \frac{1 + k_{l} \tau_c + k_{dp} \tau_c}{1 + k_{dp} \tau_c+ k_{l} \tau_c + k_{dp} k_{l} \tau_c}  \ll  1. 
\end{aligned}
\end{equation}
In a manner analogous to the reduction of the inequality in Eq. \ref{constraint_kk1_3} from Eq. \ref{constraint_kk1_2}, the above constraint simplifies to
\begin{equation}
\begin{aligned}
\bar{p} \, (1-\bar{p}) \, \tau_c \, \frac{I_{Ca}^2 \, k_{k1}^2 \, k_{dp}}{k_{d1}^2 \, k_{l} \, (k_{dp} + k_l)}  & \ll  1.
\end{aligned}
\label{constraint_kd1_2}
\end{equation}
Substituting $\bar{p}$ and $\tau_c$ from Eq.~\ref{parameters_rms} in above equation gives:
\begin{equation}
\begin{aligned}
\frac{k_+ \, \bar{c} \, k_-}{(k_+ \, \bar{c} + k_-)^3} \,  \frac{I_{Ca}^2 \, k_{k1}^2 \, k_{dp}}{k_{d1}^2 \, k_{l} \, (k_{dp} + k_l)}  & \ll  1.
\end{aligned}
\label{constraint_kd1_3}
\end{equation}
This inequality suggests that improved linearization is achieved when \(I_{Ca}\), \(k_{k1}\), and \(k_{dp}\) have small values, while larger values of \(k_{d1}\) and \(k_l\) favor better linearization.  For small values of \(k_{+} \, \bar{c}\) and \(k_{-}\), linearization improves as these parameters approach zero. On the other hand, for larger values of \(k_{+} \, \bar{c}\) and \(k_{-}\), better linearization is attained as these parameters increase.

\subsection{$k_{l} \gg \text{other cytoplasmic rate constants }$}

When \(k_{l} \gg \text{other cytoplasmic rate constants}\), the prefactor in Eq.~\ref{constraint_general} simplifies to $\frac{I_{Ca} \, k_{k1} \, \bar{p} \, (1-\bar{p})}{2 \, I_{Ca} \, k_{k1} \, \bar{p} + k_{d1} \, k_{l}} \approx \frac{I_{Ca} \, k_{k1} \, \bar{p} \, (1-\bar{p})}{k_{d1} \, k_{l}}.$ Substituting \(\frac{\Delta c_{\text{rms}}}{\bar{c}}\) from Eq.~\ref{largekl2} into Eq.~\ref{constraint_general} results in further simplification:
\begin{equation}
\begin{aligned}
\frac{I_{Ca} \, k_{k1} \, \bar{p} \, (1-\bar{p})}{k_{d1} \, k_{l}}  \, \left[ \frac{k_{dp} k_{d1} \tau_c}{\bar{p} \, (1 - \bar{p})} \, \frac{1 + k_{d1} \tau_c + k_{dp} \tau_c}{(k_{dp} + k_{d1})(1 + k_{dp} \tau_c)  (1 + k_{d1} \tau_c)} \right]^{\frac{1}{2}} & \ll 1.
\end{aligned}
\label{constraint_kl_1}
\end{equation}
This inequality acts as a constraint for the validity of Eq. 14. It is analogous to the inequality in Eq.~\ref{constraint_kd1_1}, with \(k_{d1}\) replacing \(k_{l}\). Therefore, the simplified expression, as derived for Eq.~\ref{constraint_kd1_1}, is:
\begin{equation}
\begin{aligned}
\bar{p} \, (1-\bar{p}) \, \tau_c \, \frac{I_{Ca}^2 \, k_{k1}^2 \, k_{dp}}{k_{l}^2 \, k_{d1} \, (k_{dp} + k_{d1})}  & \ll  1.
\end{aligned}
\end{equation}
Substituting $\bar{p}$ and $\tau_c$ from Eq.~\ref{parameters_rms} in above equation gives:
\begin{equation}
\begin{aligned}
\frac{k_+ \, \bar{c} \, k_-}{(k_+ \, \bar{c} + k_-)^3} \,  \frac{I_{Ca}^2 \, k_{k1}^2 \, k_{dp}}{k_{l}^2 \, k_{d1} \, (k_{dp} + k_{d1})}  & \ll  1.
\end{aligned}
\label{constraint_kl_2}
\end{equation}
This inequality indicates that better linearization is achieved when $I_{Ca}$, $k_{k1}$, and $k_{dp}$ are small, while larger values of $k_{l}$ and $k_{d1}$ enhance the linearization. Additionally, for small $k_{+} \, \bar{c}$ and $k_{-}$, linearization improves as these parameters decrease toward zero. Conversely, for larger values of $k_{+} \, \bar{c}$ and $k_{-}$, the linearization becomes more accurate as these parameters increase.

\section{}
\label{appendI}

\subsection{Time scale for cytoplasmic reactions $\gg \tau_c$}

Eq. 6 can be written as:
\begin{equation}
\begin{aligned}
\frac{\Delta c_{rms}}{\bar{c}} &=  \bigg[ \frac{\tau_c}{\bar{p} \, (1 - \bar{p})} \, \cdot d(a^2-b) \cdot \frac{4 (a + d) (1+ (a + d) \, \tau_c) + a \, [(a + 2d)^2-b] \, \tau_c^2}{a  \, [(a + 2d)^2-b] \, (1 + d \tau_c) \, [4 + 4a \tau_c + (a^2 - b) \tau_c^2]} \bigg]^{\frac{1}{2}}.
\end{aligned}
\end{equation}

When time scale for cytoplasmic reactions $\gg \tau_c$, above equation can be approximated into:
\begin{equation}
\begin{aligned}
\frac{\Delta c_{rms}}{\bar{c}} \approx  \bigg[\frac{\tau_c}{\bar{p} \, (1 - \bar{p})} \cdot \frac{d(a^2-b) \, (a + d)}{a  \, [(a + 2d)^2-b]} \bigg]^{\frac{1}{2}}.
\end{aligned}
\end{equation}
By applying the condition $d \ll a$ (refer to Eq. \ref{parameters}, where $a$ is consistently greater than $d$ unless $d=k_{dp}$ significantly exceeds the other rate parameters), above equation can be further simplified to:

\begin{equation}
\begin{aligned}
\frac{\Delta c_{rms}}{\bar{c}} \approx  \bigg[\frac{\tau_c}{\bar{p} \, (1 - \bar{p})} \cdot \frac{d(a^2-b) \, a}{a  \, (a^2-b)} \bigg]^{\frac{1}{2}} = \bigg[\frac{k_+ \, \bar{c} + k_-}{k_+ \, \bar{c} \, k_-} \cdot k_{dp} \bigg]^{\frac{1}{2}} = \sqrt{k_{dp} \, \bigg( \frac{1}{k_-} + \frac{1}{k_+ \, \bar{c}} \bigg)},
\end{aligned}
\end{equation}
which is just the Berg-Purcell result \cite{Berg1977}, given by $\frac{\Delta c_{\text{rms}}}{\bar{c}} \approx \sqrt{\frac{2 \, \tau_b}{T \, \bar{p}}} = \sqrt{\frac{2}{T} \left( \frac{1}{k_-} + \frac{1}{k_+ \, \bar{c}} \right)}$, assuming the measurement time $T$ goes as $k_{\text{dp}}^{-1}$. \\

The linearization condition specified in Eq.~\ref{constraint_general} for this case can be expressed as:
\begin{equation}
\begin{aligned}
\frac{(1-\bar{p})}{2 + \frac{k_{d1} \, k_{l}}{I_{Ca} \, k_{k1} \, \bar{p}}} \, \sqrt{k_{dp} \, \bigg( \frac{1}{k_-} + \frac{1}{k_+ \, \bar{c}} \bigg)} & \ll 1.
\end{aligned}
\end{equation}
The first fraction in the inequality above is $\frac{(1-\bar{p})}{2 + \frac{k_{d1} \, k_{l}}{I_{Ca} \, k_{k1} \, \bar{p}}} < 1$, which allows us to simplify the inequality by focusing on the second fraction. Thus, we need only ensure that the following constraint holds:
\begin{equation}
\begin{aligned}
\sqrt{k_{dp} \, \bigg( \frac{1}{k_-} + \frac{1}{k_+ \, \bar{c}} \bigg)} & \ll 1.
\end{aligned}
\end{equation}

\subsection{Time scale for cytoplasmic reactions $\ll \tau_c$}

When time scale for cytoplasmic reactions $\ll \tau_c$, Eq. 6 can be approximated into:
\begin{equation}
\begin{aligned}
\frac{\Delta c_{rms}}{\bar{c}} &\approx \bigg[ \frac{ d(a^2-b) \tau_c}{\bar{p} \, (1 - \bar{p})} \, \frac{4 (a + d) (a + d) \, \tau_c + a \, [(a + 2d)^2-b] \, \tau_c^2}{a  \, [(a + 2d)^2-b] \, d \tau_c \, [4a \tau_c + (a^2 - b) \tau_c^2]} \bigg]^{\frac{1}{2}}  \\
&= \bigg[ \frac{(a^2-b)}{\bar{p} \, (1 - \bar{p})} \, \frac{4 (a + d) (a + d) \, \tau_c + a \, [(a + 2d)^2-b] \, \tau_c^2}{a  \, [(a + 2d)^2-b] \, [4a \tau_c + (a^2 - b) \tau_c^2]} \bigg]^{\frac{1}{2}} \\
&\approx \bigg[\frac{(a^2-b)}{\bar{p} \, (1 - \bar{p})} \, \frac{a \, [(a + 2d)^2-b] \, \tau_c^2}{a  \, [(a + 2d)^2-b] \, (a^2 - b) \tau_c^2} \bigg]^{\frac{1}{2}} \quad \text{$\tau_c^2$ terms more dominating than $\tau_c$} \\
&= \bigg[\frac{1}{\bar{p} \, (1 - \bar{p})} \bigg]^{\frac{1}{2}} \\
&= \frac{k_+ \, \bar{c} + k_-}{\sqrt{k_+ \, \bar{c} \cdot k_-}}.
\end{aligned}
\label{error_I}
\end{equation}
This result is independent of the cytoplasmic rate constants, similar to the Berg-Purcell result. However, unlike the Berg-Purcell prediction, the error does not necessarily decrease with increasing ligand binding and unbinding rates. Instead, the error may increase. For example, when \( k_+ \, \bar{c} \gg k_- \), above equation can be approximated as \( \frac{\Delta c_{\text{rms}}}{\bar{c}} \approx \sqrt{\frac{k_+ \, \bar{c}}{k_-}} \), whereas when \( k_+ \, \bar{c} \ll k_- \), it simplifies to \( \frac{\Delta c_{\text{rms}}}{\bar{c}} \approx \sqrt{\frac{k_-}{k_+ \, \bar{c}}} \).  \\
The linearization condition specified in Eq.~\ref{constraint_general} for this case is expressed as:
\begin{equation}
\begin{aligned}
\frac{(1-\bar{p})}{2 + \frac{k_{d1} \, k_{l}}{I_{Ca} \, k_{k1} \, \bar{p}}} \, \frac{k_+ \, \bar{c} + k_-}{\sqrt{k_+ \, \bar{c} \cdot k_-}} & \ll 1.
\end{aligned}
\label{linearization_I}
\end{equation}
As previously noted, the first fraction in the inequality above is $\frac{(1-\bar{p})}{2 + \frac{k_{d1} \, k_{l}}{I_{Ca} \, k_{k1} \, \bar{p}}} < 1$. Thus, we need only ensure that the following constraint holds:
\begin{equation}
\begin{aligned}
\frac{k_+ \, \bar{c} + k_-}{\sqrt{k_+ \, \bar{c} \cdot k_-}} & \ll 1.
\end{aligned}
\label{eq:I8}
\end{equation}

For \( k_+ \, \bar{c} = k_- \), the error (given by Eq. \ref{error_I}) reaches its minimum value of \(2\), indicating that when the timescale of cytoplasmic reactions is much smaller than \(\tau_c\) and the attachment rate equals the detachment rate, the error becomes independent of both ligand-receptor kinetics and cytoplasmic rates. In this case, linearization constraint from Eq.~\ref{linearization_I} can be expressed as:
\begin{equation}
\frac{0.5}{2 + \frac{k_{d1} \, k_{l}}{I_{Ca} \, k_{k1} \, 0.5}} \cdot 2 \ll 1 \quad \Rightarrow 2 + \frac{2 \, k_{d1} \, k_{l}}{I_{Ca} \, k_{k1}} \gg 1.
\end{equation}
As \( 2 > 1 \). Thus, we need only ensure that the following constraint holds:
\begin{equation}
\frac{2 \, k_{d1} \, k_{l}}{I_{Ca} \, k_{k1}} \gg 1.
\end{equation}

\section{Sensitivity to Gradients in \textit{Dictyostelium}}
\label{appendJ}

Consider a \textit{Dictyostelium} cell, where the number of occupied receptors on the front half, back half, and total surface of the cell are denoted by \(N_f\), \(N_b\), and \(N\), respectively, under threshold concentration and spatial gradient sensing conditions. The concentration difference between the front and back halves is denoted by \( (\Delta c)_{\text{across cell}} = c_f - c_b \), where \(c_f\) and \(c_b\) represent the concentrations at the front and back ends, respectively. This difference has to be detected by Dictyostelium, which operates in the presence of a background concentration of cyclic AMP (\(c\)). The ratio $(\Delta c)_{\text{across cell}}/c$ serves as a measure of the Dictyostelium's sensitivity to gradient \cite{Mato1975}. Using error propagation for \( (\Delta c)_{\text{across cell}} = \Delta c = c_f - c_b \), the following relation can be written:
\begin{equation}
\begin{aligned}
\frac{\sigma_{\Delta c}^2}{c^2} = \frac{\sigma_{c_f}^2}{c^2} + \frac{\sigma_{c_b}^2}{c^2} = \frac{1}{N_f} \, \frac{\sigma_{c}^2}{c^2} + \frac{1}{N_b} \, \frac{\sigma_{c}^2}{c^2} = \frac{N_f + N_b}{N_f \, N_b} \, \frac{\sigma_{c}^2}{c^2},
\end{aligned} 
\end{equation}
where \(\frac{\sigma_{c}}{c} = \frac{\Delta c_{\text{rms}}}{\bar{c}}\) represents the RMS error in sensing the ligand concentration using a single receptor (as given by Eq.~6 in the main text). Using \(N_f - N_b = \Delta N_{fb}\) (the difference in the number of occupied receptors between the front and back halves of the cell at threshold sensing) and \(N_f + N_b = N\), we obtain the expressions \(N_f = \frac{N + \Delta N_{fb}}{2}\) and \(N_b = \frac{N - \Delta N_{fb}}{2}\). Thus, the above equation simplifies to:
\begin{equation}
\begin{aligned}
\frac{\sigma_{\Delta c}}{c} = \sqrt{ \frac{4N}{(N + \Delta N_{fb})(N - \Delta N_{fb})} } 
\, \cdot \, \frac{\Delta c_{\text{rms}}}{\bar{c}}
\end{aligned} 
\label{gradient_sensitivity}
\end{equation}
The result for sensitivity to the gradient in \textit{Dictyostelium} cells is calculated using above equation.

\section{Simulation Methodology}
\label{appendK}

In our simulation, we employ the Gillespie algorithm to model receptor activation. The Gillespie algorithm starts at time \( t \) and computes the time to the next receptor activation event using the formula  

\[
\Delta t = -\frac{\log(r_1)}{k_+}
\]

where \( r_1 \) is a random number uniformly distributed between 0 and 1, and \( k_+ \) is the activation rate constant. For receptor deactivation, the algorithm calculates the time until the next deactivation event using  

\[
\Delta t = -\frac{\log(r_2)}{k_-}
\]

where \( r_2 \) is another uniformly distributed random number, and \( k_- \) represents the deactivation rate constant. This process continues iteratively until the simulation reaches the predefined measurement time \( T \).  

The remaining part of the rate equations, excluding the stochastic calcium influx \( I_{Ca} \cdot p(t) \), is solved deterministically using an explicit Euler method to update the concentration profiles of all species. At measurement time \( t \), this process yields the phosphorylation readout \( C_{Pr}(t) \). The mean phosphorylation readout \( \bar{C}_{Pr} \) and the variance in phosphorylation readout \( \sigma_{C_{Pr}}^2 \) are calculated by using the \( N \) values of \( C_{Pr}(t) \), where \( N \) represents the number of times the simulation is repeated. The RMS error in inferring the ligand concentration is obtained as  

\[
\frac{\Delta c_{rms}}{\bar{c}} = \frac{\sigma_{C_{Pr}}}{\bar{c} \, \frac{\Delta \bar{C}_{Pr}}{\Delta \bar{c}}}
\]

where \( \bar{c} = c \) represents the ligand concentration that the cell needs to sense, and a slight change in this concentration, \( \Delta \bar{c} \), is introduced to calculate the resulting change in the mean phosphorylation readout \( \Delta \bar{C}_{Pr} \).

% Bibliography

\end{document}